\documentclass[12pt]{article}

\usepackage{amsmath,amsfonts,latexsym,amssymb,amscd,amsbsy,dsfont}
\usepackage{pslatex}
\usepackage[latin1]{inputenc}
\usepackage[T1]{fontenc}
\usepackage{pspicture}
\usepackage{verbatim,amsthm,curves,graphics}
\usepackage{mathrsfs}
\usepackage[dvips]{graphicx}
\usepackage{pstricks}
\usepackage{color}

\usepackage{tikz}
\usetikzlibrary{shapes.geometric}
\usetikzlibrary{decorations.markings}
\usetikzlibrary{decorations.pathmorphing}
\def\f(#1){{\mathop{f}^{(#1)}}}
\def\m(#1){{\mathop{m}^{(#1)}}}
\def\C(#1){{\mathop{C}^{(#1)}}}
\def\p(#1){{\mathop{p}^{(#1)}}}

\def\ben{\begin{equation}}
\def\een{\end{equation}}
\def\bena{\begin{eqnarray}}
\def\eena{\end{eqnarray}}
\def\non{\nonumber}

\def\V{{\cal V}}

\def\I{{\mathfrak I}}
\def\R{{\mathfrak R}}
\def\V{{\mathcal V}}
\def\N{{\cal N}}
\def\D{D}
\def\mr{{\mathbb R}}

\def\M{{M}}

\newcommand{\mn}{{\mathbb N}}
\newcommand{\mz}{{\mathbb Z}}
\newcommand{\cst}{{\rm cst.}}

\newcommand{\mc}{{\mathbb C}}

\renewcommand{\H}{{\mathcal H}}

\newcommand{\hF}{{}_2 F_1}
\renewcommand{\D}{{\mathbb D }}
\renewcommand{\H}{{\mathbb H }}

\renewcommand{\C}{{\mathcal C}}

\newcommand{\ctg}{\operatorname*{ctg}}

\newcommand{\E}{{\mathcal E}}
\newcommand{\e}{{\rm e}}
\renewcommand{\p}{{\bf p}}
\newcommand{\x}{{\bf x}}
\newcommand{\q}{{\bf q}}

\theoremstyle{definition}

\newtheorem{thm}{Theorem}

\newtheorem{prop}{Proposition}

\begin{document}

\title{Massless interacting quantum fields in deSitter spacetime}

\author{
Stefan Hollands$^{1}$\thanks{\tt HollandsS@Cardiff.ac.uk}\:,
\\ \\
{\it ${}^{1}$School of Mathematics,
     Cardiff University, UK}
}
\date{20 May 2011}
\maketitle

\begin{abstract}
We generalize key aspects of gr-qc/1010.5364 (and also gr-qc/1010.5327)
to the case of {\em massless} $\lambda \phi^{2n}$ quantum field theory
on deSitter spacetime. As in that paper, our key objective is to derive a suitable ``Mellin-Barnes-type''
representation of deSitter correlation functions in a deSitter-invariant state, which holds to arbitrary orders in perturbation theory, and which incorporates renormalization. The representation is suitable
for the study of large distance/time properties of correlation functions. It
is arrived at via an analytic continuation from the corresponding objects on the sphere, and, as in the massive case, relies on the use of graph-polynomials and their properties, as well as other tools. However, the perturbation expansion is organized somewhat differently in the massless case, due to the well-known subtleties associated with the ``zero-mode'' of the quantum field. In particular, the correlation functions do not possess a well-defined limit as the self-coupling constant of the field goes to zero, reflecting the well-known non-existence of a deSitter invariant state in the free massless scalar theory. We establish that generic correlation functions cannot grow more than polynomially in proper time for large time-like separations of the points. Our results thus leave open
the possibility of quantum induced IR-instabilities of deSitter spacetime on very large time-scales.
\end{abstract}

\pagebreak

\tableofcontents

\pagebreak


\sloppy

\section{Introduction}

Perturbative calculations of quantum field theory correlators in expanding
FRW-spacetimes play an important role in cosmology, e.g. for the quantitative understanding
of the finer details of the formation of density perturbations from quantum fluctuations
in the Early Universe. Specifically, one is interested in this context mostly in correlation functions
of the type
\ben\label{cosmobs}
\langle \phi(t, {\bf p}_1) \cdots \phi(t, {\bf p}_E) \rangle_\Psi \ ,
\een
where $\phi$ is some quantum field---typically related to the inflaton---and
where ${\bf p}_i$ are spatial momenta associated with a slice
in a flat FRW-spacetime $ds^2 = -dt^2 + a(t)^2 d{\bf x}^2$. $t$ is
a suitably chosen time, taken usually when all physical wave numbers ${\bf p}_i/a(t)$ are
considerably smaller in magnitude than the Hubble rate $\dot a(t)/a(t)$. For
the power spectrum in the CMB, $E=2$ is relevant, for ``non-Gaussianities'' $E=3$ or higher.
The state $\Psi$ in question is typically fixed as an ``in-state'' in the very
Early Universe. It is of considerable interest to:
\begin{enumerate}
\item Develop methods to calculate such correlation functions systematically in
renormalized perturbation theory.
\item Understand the extent to which the precise choice of the in-state affects the result,
and---a closely related question---to investigate the IR-behavior of correlators for large times.
\end{enumerate}
A spacetime which is particularly relevant in inflationary cosmology is deSitter space,
with scale factor $a(t) = \e^{tH}$. This spacetime has the additional merit of
possessing the same number of continuous symmetries as Minkowski spacetime.
One may therefore hope to be able to develop
a formalism for perturbative calculations of comparable efficiency and elegance as in Minkowski spacetime.

In~\cite{self} and~\cite{marolf2}, significant progress in these directions was made
for the case of {\em massive} scalar field theories on deSitter spacetime. In these papers
``Mellin-Barnes''-type representations were derived for an arbitrary Feynman graph contributing to an $E$-point correlation function in a special deSitter invariant state. Such representations
express a correlation function in terms of complex powers the deSitter invariants formed from the external points
in position space. The powers are integrated over contours in the complex plane, thus providing
an explicit expression for the renormalized correlators of the desired type 1). The momentum space
correlators can be obtained from these by a further Fourier transformation.  The representation
allows one to see that correlation functions have an exponential decay in position space in
time-like directions, such as e.g.
\ben
\langle \phi(X_1) \phi(X_2) \rangle_0 = O(\e^{-\cst H\tau}) \ ,
\een
where $X_1, X_2$ are deSitter points separated by a large proper time $\tau$, and
where ``0'' denotes the deSitter invariant state. Similar
results were proved in~\cite{self,marolf2} also for arbitrary $E$-point functions, to arbitrary orders in
perturbation theory, see also~\cite{marolf3,higuchi1}. The positive constant is related to
to the mass of the theory. Estimates of this form imply a quantum version of the ``no-hair theorem''.
This states that the correlation functions in essentially {\em any}
state will approach those of the deSitter invariant state  at late times. In particular, expectation values of operators in an essentially arbitrary state will approach those of
the deSitter invariant state at late times. Hence, this analysis in particular answered question 2) above.
For previous contributions to this issue, including other claims, see e.g.~\cite{mottola,higuchi,polyakov,tanaka,weinberg1,weinberg2,weinberg3,woodard}.

Both papers~\cite{self,marolf2} left open the case of a massless theory, such
as
\ben\label{lagr}
L = \left[\frac{1}{2}(\nabla \phi)^2 + \lambda \phi^{2n} \right] \ d\mu \ , \ \ \ \quad \lambda > 0 \ .
\een
The purpose of this paper is to close this gap and to address questions 1), 2) for such
theories. With respect to 1), we obtain formulae of a similar type for the $E$-point correlation
functions as in the
massive theory, see eq.~\eqref{gk2}, or the related representation given in appendix~\ref{app:D}. The state in which these correlation functions are computed is a deSitter invariant state,
which is obtained via analytic continuation from the sphere. Unlike in the massive theory,
this state has no limit as the coupling $\lambda \to 0$, as indeed, there is no such state
in the free theory~\cite{allen}. With regard to 2), we find that the exponential decay expectedly no longer holds in the massless
theory, but we can show that the correlation functions {\em grow} no faster than a polynomial
in $H\tau$. Unfortunately the order of the polynomial can depend on the perturbative order.
We believe that it should be possible to obtain that the correlator actually {\em decays}
as a polynomial in $H\tau$, and we present in appendix~\ref{app:B} some evidence for the two-point function based on the so-called ``K\" allen-Lehmann representation'' in deSitter space~\cite{bros}. Our results imply that,
in practice, correlation functions in arbitrary states are expected to decay/increase only over cosmolgical time-scales of the order of the Hubble time.

The feature of massless deSitter scalar fields that complicates our analysis is that--as we have mentioned--the
underlying free ($\lambda =0$) field theory does not have a deSitter invariant state,
unlike massive theory. In the Euclidean counterpart of the theory on a sphere, this
shows up through the presence of a zero mode in the massless case. In order to find a perturbation expansion of the correlators of a deSitter
state in the interacting theory, one can use a modification of the standard perturbative setup~\cite{raja} which takes into account the zero mode, as is also often done for perturbations of
massless fields in 2-dimensional flat space.

The difference
between the massive/massless case can already be
seen, roughly speaking, from the behavior of the 2-point function, $W(m^2; X_1, X_2)$, of
the deSitter invariant state in
the free theory, where $m^2>0$ is the mass parameter. As we will recall in the paper,
this 2-point function is given by a hypergeometric function, whose decay for large time-separation
is $\sim \e^{-\cst H \tau}$. The positive dimensionless constant is of order
$\cst=O(m^2)$ for $m^2 \to 0$, so the decay gets weaker.
In the free theory, one cannot set $m^2 = 0$. But in an interacting theory~\eqref{lagr}, one can set $m^2 = 0$ as long as $\lambda>0$.  For an interacting theory, we argue in appendix~\ref{app:B}
that the 2-point function has the ``K\" allen-Lehmann representation''
\ben
\langle \phi(X_1) \phi(X_2) \rangle_0 = \int_0^\infty dM^2 \ \rho(M^2) \ W(M^2; X_1, X_2) \ .
\een
[In appendix~\ref{app:B}, we also outline an algorithm how to calculate the weight $\rho(M^2)$ in perturbation theory.]
A massive theory is characterized by the fact that $\rho(M^2)$ has its support contained in
$[m^2, \infty)$ for some $m^2 > 0$. By contrast, in the massless theory,
the support contains the point $M^2=0$.
Because the decay of $W(M^2; X_1, X_2)$ in time-like directions is only of order $\e^{-O(M^2) \tau}$
for small $M^2$, it is intuitively clear that the full two-point function in the massless
theory cannot decay exponentially as it does in the massive theory.

\medskip

\noindent
{\em Notations:} The dimension of spacetime is $D \ge 2$, and is always integer. $\E G$ denotes the set of edges of a Feynman graph $G$, and
$\V G$ the set of vertices. $\R(z),\I(z)$
denote the real and imaginary part of a complex number. $d\mu$ is the invariant
integration measure on the $D$-sphere $S^D$, resp. real deSitter space $dS_D$,
depending on the context. Capital letters such as $X$ denote points in
$S^D$ resp. $dS_D$ (depending on the context), viewed as embedded hypersurfaces in
$\mr^{D+1}$. $X \cdot X$ etc. denotes the Euclidean resp. Minkowskian inner product
in $\mr^{D+1}$, depending on the context. For the rest of this paper, except in section~\ref{subsec:phys},
we will set
\ben
H = 1 \ .
\een
However, $H$ can be re-introduced in all of our formulas by simple dimensional analysis: We have $[X] = 1, [\phi] = -D/2+1, [m]=-1, [\lambda]=-D+2n(D/2-1), [Z]=0, [\tau] = 1, [d\mu]=D$ for the length dimensions.
Appropriate powers of $H$ then have to be inserted to match the dimensions, with
$[H]=-1$. For example, to make the point-pair invariant $Z$ dimensionless [cf. eq.~\eqref{pointpair}], we should set $Z = H^2 X_1 \cdot X_2$, the free-field covariance [cf. eq.~\eqref{Cdef}] should have dimension $[C] = 2[\phi]$, so on the right side we should multiply by $H^{D-2}$, etc.

\section{Massless Euclidean quantum field theory on $S^D$}

\subsection{Euclidean path integral}

Correlation functions in a Euclidean ``quantum field theory'' on the $D$-dimensional sphere, $S^D$, are formally
defined by a path integral
\ben
\langle \phi(X_1) \dots \phi(X_E) \rangle_0 = \frac{1}{\N} \int D\phi \ \e^{-I} \ \phi(X_1) \dots \phi(X_E) \ .
\een
Here $I$ is an action. In our case it is of the form
\ben
I = \int_{S^D} \left[ \frac{1}{2} (\nabla \phi)^2 + P(\phi) \right] d\mu
\een
with some polynomial potential $P(\phi)$. The path integral has no chance to make sense unless
$P(\phi) \ge \cst$, i.e. unless the interaction is stable. The example to which we will stick
from now on is
\ben
P(\phi) = \frac{1}{2} m^2 \phi^2 + \lambda \phi^{2n} \  \ .
\een
For $m^2 \ge 0$, stability requires that $\lambda \ge 0$ (with $\lambda = 0$ allowed).
For $m^2 < 0$, stability requires that $\lambda > 0$ (with $\lambda = 0$ not allowed).
Even if these condition are satisfied,
the path integral remains only a symbolic expression at best, because the
infinite dimensional Lebesgue measure $D\phi$ does not exist. It requires a lot of effort
to make sense of the path integral, and this has been achieved in fact only for $D=2,3$
(and $n=2$ in $D=3$), and mostly in flat space, see e.g.~\cite{glimm}. In $D=4$, the path integral probably does not exist unless $\lambda = 0$ (``triviality problem''), meaning that the theory probably has no UV-completion in
the non-perturbative setting.

At any rate, in practice, one cannot perform the path integral exactly. Instead, it is normally interpreted
as the formal power series
\ben\label{pert}
\langle \phi(X_1) \dots \phi(X_E) \rangle_0 = \frac{1}{\N} \sum_{V=0}^\infty \frac{(-\lambda)^V}{V!} \int D\phi \ \e^{-I_0} \
\phi(X_1) \dots \phi(X_E)
\left( \int_{S^D} \phi^{2n}(Y) \ d\mu(Y)
\right)^V \ .
\een
In this expression, $I_0$ is the free action, given by
\ben
I_0 = \int_{S^D} \left( \frac{1}{2} (\nabla \phi)^2 + \frac{1}{2} m^2 \phi^2 \right) d\mu \ .
\een
While in the exact path integral, the value of $m^2$ was arbitrary as long as $\lambda > 0$, this
is now no longer so for the individual terms in the perturbation expansion. In this paper, we
are interested in the massless case, which is
\ben
m^2 = 0 \ .
\een
This choice creates a problem in the ``naive'' perturbation expansion~\eqref{pert}, as
can already be seen at the first term $V=0$. Then, e.g. for
$E=2$, we formally only have to do a Gaussian integration, namely (up to a constant)
\ben
\int D\phi \ \e^{-I_0} \ \phi(X_1) \phi(X_2) = \left( \frac{1}{-\nabla^2} \right)(X_1, X_2) \, .
\een
Here, $\nabla^2$ is the Laplacian on $S^D$. Unfortunately, this operator has a kernel--the constant functions, or
``zero modes''--so its inverse actually does not exist! Thus, already the first term in the perturbation expansion does not
exist. This reflects the well-known obstruction to define a ``Euclidean vacuum state'' for a massless scalar field. The same problem also occurs for higher orders in the naive perturbation expansion around such a field.

To see this somewhat more systematically, let us decompose the field $\phi(X)$ into the contribution from
the zero mode called $x$, and the rest called $\psi(X)$:
\ben
\phi(X) = x + \psi(X) \ , \quad \psi(X) = \sum_{L=1}^\infty \sum_{m=1}^{N(D,L)} \phi_{Lm} \ Y_{Lm}(X) \ ,
\een
where $Y_{Lm}, L = 0,1,2,..., m=1,...,N(D,L)$ are the spherical harmonics on $S^D$ (see appendix~\ref{app:appa}) satisfying $-\nabla^2 Y_{Lm} = L(L+D-1) \ Y_{Lm}$. The Gaussian measure then formally becomes, up to a constant
\ben
\int D\phi \ \e^{-I_0} = \int_{-\infty}^{+\infty} dx \int d\nu_C(\psi) \ ,  
\een
where $d\nu_C(\psi)$ is the Gaussian measure of ``covariance $C$''. By definition
\ben\label{wick}
\int d\nu_C(\psi) \ \prod_{j=1}^n \psi(X_j) =
\begin{cases}
0 & \text{if $n$ is odd,} \\
\prod_{\text{pairs} \ (ij)} C(X_i, X_j) & \text{if $n$ is even} \  .
\end{cases}
\een
The covariance $C$ is the propagator for $\psi$ and is
 given by the ``Green's function of $-\nabla^2$ minus the zero mode'', i.e.
\ben\label{Cdef1}
C(X_1, X_2) = \sum_{L=1}^\infty \sum_{m=1}^{N(D,L)} \frac{Y_{Lm}(X_1)^* Y_{Lm}(X_2)}{L(L+D-1)} \ .
\een
The problem with the naive perturbation expansion~\eqref{pert} is now manifest: The integration $\int_{-\infty}^{+\infty} dx$
implicit in each term in eq.~\eqref{pert} is not convergent, because the integrand is a polynomial in $x$.
Of course, this problem was not present in the original path integral, where the exponential provided a strong
damping for large $|x|$. Thus, to remedy the problem, one has to abolish the naive perturbation expansion~\eqref{pert}. However, as suggested by~\cite{raja}, rather than going back to the full non-perturbative (not expanded) path integral, which is untractable, one can steer a middle path and consider only the $x$-integration non-perturbatively, while expanding the rest out. This ``semi-perturbative'' expansion looks in more detail as follows.

First, we write the potential term as
\bena\label{pdef}
P(\phi(X)) &=& \lambda[x + \psi(X)]^{2n} \non\\
&=& \lambda x^{2n} + \lambda \sum_{k=0}^{2n-1} \left(
{2n \atop k}
\right) \ x^k \ \psi^{2n-k}(X) \non\\
&=:& \lambda x^{2n} + \lambda p(x, \psi(X)) \ .
\eena
To shorten some formulas, it is then useful to redefine
\ben
\lambda \to {\rm vol}(S^D)^{-1} \ \lambda \ .
\een
We now expand the weighted measure in the path integral as
\bena
D\phi \ \e^{-I} &=& D\phi \ \e^{-I_0} \ \e^{-\lambda x^{2n}} \ \sum_{V=0}^\infty \frac{1}{V!} \left(\frac{-\lambda}{{\rm vol}(S^D)}
\right)^V  \left(
\int_{S^D} p(x, \psi) d\mu \right)^V \non\\
&=& dx  \ d\nu_C(\psi)  \ \e^{-\lambda x^{2n}} \ \sum_{V=0}^\infty \frac{1}{V!}
\left(\frac{-\lambda}{{\rm vol}(S^D)} \right)^V \left(
\int_{S^D} p(x, \psi) d\mu \right)^V \ .
\eena
We insert this into the path integral, and we also replace $\phi(X_j) = x + \psi(X_j)$. The result is
\bena
\langle \phi(X_1) \dots \phi(X_E) \rangle_0 &=& \frac{1}{\N} \sum_{V=0}^\infty \frac{1}{V!}
\left(\frac{-\lambda}{{\rm vol}(S^D)} \right)^V
\int_{-\infty}^\infty dx  \ \e^{-\lambda x^{2n}} \int d\nu_C(\psi)   \ \left( \int_{S^D} \prod_{i=E+1}^{V+E} d \mu(X_i) \right)  \non\\
&& \vspace{2cm} \times \ \prod_{i=1}^E \ [x + \psi(X_i)] \ \prod_{j=E+1}^{V+E} p(x, \psi(X_j)) \ .
\eena
Note that the potential $p(x, \psi)$ is polynomial in $x$ (and also $\psi$). Hence, the $dx$-integral
can be performed explicitly using the standard formula
\ben\label{dxi}
\int_{-\infty}^{\infty} dx \ \e^{-\lambda x^{2n}} \ x^{s-1} = \frac{1}{n} \lambda^{-s/2n} \Gamma(s/2n) \ ,
\een
for $s$ odd, and $=0$ for $s$ even.
The $d\nu_C(\psi)$ functional integral is then performed (formally) using the
rule eq.~\eqref{wick} for Gaussian integration. The covariance (``propagator'' of $\psi$) is $C(X_1,X_2)$,
rather than the non-existent Green's function of $\nabla^2$. We hence get an expansion in terms of Feynman
graphs. Each vertex in such a graph $G$ corresponds to an interaction term in $p(x, \psi)$,
whereas the propagators are given by $C$.

Carrying out these straightforward manipulations results in the following expansion:
\bena\label{ig0}
\langle \phi(X_1) \dots \phi(X_E) \rangle_0 &=& \frac{1}{\N} \ \sum_{\text{graphs $G$}}
c_G \ \lambda^{a_G} \ I_G(X_1, \dots, X_E) \ .
\eena
The sum is over all Feynman graphs $G$ with $V$ internal vertices, and
$E$ external vertices labeled by $X_1, \dots, X_E$. The valence of the
internal vertices is any number from $\{2,3, ..., 2n\}$, whereas the
valence of the external vertices is a number from $\{0,1\}$. The power $a_G$ of the coupling
constant differs from the usual power $\lambda^V$, and is due to the $dx$-integrations~\eqref{dxi}. It no longer
has to be non-negative, nor integer, but is given by
\ben
a_G = \frac{1}{2n} \sum_j k_j - \frac{E}{2n} \ ,
\een
where $k_j$ is the incidence number of the $j$-th vertex of the graph $G$.
$c_G$ is a constant depending on the combinatorial structure of the graph.
It is given by
\ben\label{cG}
c_G = \left[\frac{(2n)!}{{\rm vol}(S^D)}\right]^V \ |{\rm Aut}(G)|^{-1}  \ \frac{\Gamma(V-\frac{1}{2n} \sum_j k_j + \frac{E}{2n})}{\prod_j (2n-k_j)!} \ .
\een
The second factor comes from the symmetry group of the graph. The third factor
comes from the $dx$-integration~\eqref{dxi}, and from coefficients in the interaction polynomial $p$, see eq.~\eqref{pdef}.
 $I_G$ is a Feynman integral. It is given by
\ben\label{igdef}
I_G(X_1, \dots, X_E) = \left( \prod_{i=E+1}^{V+E} \ \int_{S^D}  d \mu(X_i) \right)
\prod_{\ell \in \E G} C(X_{s(\ell)}, X_{t(\ell)}) \ ,
\een
where the product is over all edges $\ell$ of the graph $G$, whose source resp. target
vertex is denoted by $s(\ell)$ resp. $t(\ell)$.
As usual, this expression is still formal because the integral has UV-singularities. These
must be dealt with using renormalization theory.

The constant $\N$ in eq.~\eqref{ig0} is defined so that $\langle 1 \rangle_0 = 1$. In standard perturbation theory,
$\N^{-1}$ simply removes the ``vacuum bubble'' diagrams $G$, and consequently does not have to be calculated. However,
because the integration over $x$ is non-Gaussian in our situation, $\N^{-1}$ now no longer just
removes the bubble diagrams. It has to be calculated explicitly. In our normalizations,
$\N = 1 + O(\lambda^{2/n})$.

\subsection{Parametric representation}\label{parametric}

We now look in more detail at the Feynman integral $I_G$, defined formally by
\eqref{igdef}. To get a better understanding of this object, we first give a
more useful representation of the covariance $C$ defined previously in~\eqref{Cdef1}. This
representation is derived in
appendix~\ref{app:appa}. It can be stated most straightforwardly if we regard points $X \in S^D$
as unit vectors in $\mr^{D+1}$, i.e. we think of $S^D$ in the following as the embedded
hypersurface
\ben
S^D = \{ X \in \mr^{D+1} \ : \ X \cdot X = X_0^2 + \dots + X_D^2 = 1 \} \ .
\een
With this interpretation, we write $(X_1-X_2)^2 = (X_1-X_2) \cdot (X_1 - X_2) = 2-2X_1 \cdot X_2$, etc. The
formula for $C$ that we will use is:
\ben
C(X_1, X_2) = \frac{h_D}{{\rm vol}(S^D)} + \frac{1}{(-4\pi)^{D/2+1/2}} \int_c dz
 \ \frac{\Gamma(D-1+z)\Gamma(-D/2+1-z)\Gamma(z)}{\Gamma(2z+1)\Gamma(-z+1/2)}
\ [(X_1-X_2)^2]^z\non \ .
\een
Here, the contour $c$ goes around the poles of the integrand
at $z=-D/2+1, ..., -1, 0$, and $h_D=\sum_{n=1}^{D-1} \frac{1}{n}$ are the harmonic numbers.
In the following, we will view $c$ as the union of circles $c_k$
given by
\ben\label{ck}
c_k: \ \ \ t \mapsto k + \epsilon \e^{2\pi i t} \ , \ \ \ \ \ \epsilon>0.
\een
We can substitute this formula for each propagator $C$ into our expression~\eqref{igdef} for the Feynman
integral. To get somewhat cleaner looking expressions in the following, it is convenient to
absorb the constant term and the pre-factor in $C$ into a redefinition in eq.~\eqref{ig0} of $I_G$ to $\tilde I_G$, of
$\N$ to $\tilde \N$, and of the combinatorial factor $c_G$ in eq.~\eqref{cG} to $\tilde c_G$ given by~\cite{tildecG}.
Then $\langle \prod \phi \rangle_0 = \tilde \N^{-1} \sum \lambda^{a_G} \ \tilde c_G \ \tilde I_G$, with
\bena\label{ig}
\tilde I_{G}(X_1, \dots, X_E) &=&  \ \bigg(  \prod_{\ell \in \E G} \int_c dz_\ell \bigg) \
\bigg( \prod_{i=E+1}^{V+E} \ \int_{S^D}  d \mu(X_i) \bigg)
\ \prod_{i,j=1}^{V+E} \left[ (X_i-X_j)^2 \right]^{z_{ij}}  \non \\
&&\times \  \prod_{\ell  \in \E G}
\frac{\Gamma(D-1+z_\ell)\Gamma(-D/2+1-z_\ell) \Gamma(z_\ell)}{ \Gamma(2 z_\ell + 1) \Gamma(-z_\ell + 1/2)} \ .
\eena
In this expression, $i,j$ is a pair of vertices, and for each such pair we have set
\ben\label{zij}
z_{ij} = \sum_{\ell \in \E G: \ell = (ij)} z_\ell \ ,
\een
i.e. $z_{ij}$ is the sum of all the parameters associated with lines that connect a given pair of
vertices $i,j = 1, \dots, V+E$. If there are no lines in the graph $G$ connecting $i,j$, then
the corresponding factor in eq.~\eqref{ig} is understood to be absent. As usual, we
assume that the external points $X_i, i=1,\dots,E$ are pairwise distinct.

The integral $\tilde I_G$ involves contour integrations, as well as integrals over $X_i \in S^D, i=E+1,...,E+V$.
These {\em `master integrals'} are
\ben\label{feyn}
M_G(X_1, ..., X_E) :=
\bigg( \prod_{i=E+1}^{V+E} \ \int_{S^D}  d \mu(X_i) \bigg)
\ \prod_{i,j=1}^{V+E} \left[ (X_i-X_j)^2 \right]^{z_{ij}} \ .
\een
They are absolutely convergent for $D=2$ and the configurations $z_{ij} \in \mc$ that we need. But for $D>2$,
the integrand is in general too singular at coinciding points $X_i = X_j$ to be integrable.
These are of course the familiar UV-divergences in perturbative quantum field theory.
To treat them, one has to use renormalization theory. This was carried out in~\cite{self} by
a method based on~\cite{hollandswald1,hollandswald2,freddue,brunetti,holfred}. The point is that the integrals over the $X_j$ are
well-defined (absolutely convergent) in any dimension provided that all $\R(z_{ij})$ are sufficiently large.
Then, it was shown that the resulting function can be analytically continued to other values
of the $z_{ij}$ provided that a certain ``absence of resonance'' condition holds between
the real parts $\R(z_{ij})$. The condition states that there should exist no integer
linear combination $\sum n_{ij} \ \R(z_{ij}) \in \mz$.
The subsequent $dz_\ell$-integration contours $c$ in $\tilde I_G$
\bena\label{ig1}
\tilde I_{G}(X_1, \dots, X_E) &=& \ \bigg(  \prod_{\ell \in \E G} \int_c dz_\ell \bigg) \
M_G(X_1, \dots, X_E; \{z_{\ell}\})  \non \\
&&\times \  \prod_{\ell  \in \E G}
\frac{\Gamma(D-1+z_\ell)\Gamma(-D/2+1-z_\ell) \Gamma(z_\ell)}{ \Gamma(2 z_\ell + 1) \Gamma(-z_\ell + 1/2)}
\eena
must then be appropriately chosen so that the absence of resonance condition is satisfied along the contour. This can e.g. be achieved by breaking each $c$ up into small circles $c_k(t)= k + \epsilon_\ell \e^{2\pi i t}$ around $k=0,-1,...,-D/2+1$, with a suitably chosen $\epsilon_\ell>0$ for each
$dz_\ell$-integration.

\subsection{Master integrals $M_G$}

\subsubsection{Example graph}

These issues can be illustrated most easily in the most straightforward example, the ``star graph'' $G$ shown in the following
 picture.
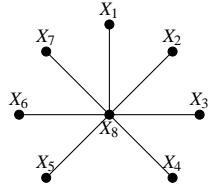
\begin{center}
\begin{tikzpicture}[scale=.6, transform shape]
\draw (0,0) node[draw,shape=circle,scale=0.5,fill=black]{};
 \draw (0,2) node[draw,shape=circle,scale=0.5,fill=black]{};
 \draw (1.4,1.4) node[draw,shape=circle,scale=0.5,fill=black]{};
 \draw (2,0) node[draw,shape=circle,scale=0.5,fill=black]{};
 \draw (1.4,-1.4) node[draw,shape=circle,scale=0.5,fill=black]{};
  \draw (-1.4,-1.4) node[draw,shape=circle,scale=0.5,fill=black]{};
  \draw (-2,0) node[draw,shape=circle,scale=0.5,fill=black]{};
   \draw (-1.4,1.4) node[draw,shape=circle,scale=0.5,fill=black]{};
 \draw[black]  (0,0) node[black,below]{$X_8$} --  (0,2) node[black,above]{$X_1$};
 \draw[black]  (0,0) --  (1.4,1.4) node[black,above]{$X_2$};
 \draw[black]  (0,0) --  (2,0) node[black,above]{$X_3$};
 \draw[black]  (0,0) --  (1.4,-1.4) node[black,above]{$X_4$};
 \draw[black]  (0,0) --  (-1.4,-1.4) node[black,above]{$X_5$};
 \draw[black]  (0,0) --  (-2,0) node[black,above]{$X_6$};
 \draw[black]  (0,0) --  (-1.4,1.4) node[black,above]{$X_7$};
\end{tikzpicture}
\end{center}
In the graph in the picture, the integration variable is $X_8$, and the external legs
are $X_1,...,X_7$, i.e. $E=7,V=1$.
The corresponding master integral~\eqref{feyn} for this graph $G$ was computed in \cite{self,marolf2} as:
\bena\label{master0}
&&\M_G(X_1, \dots, X_E) = \frac{(4\pi)^{D/2+1/2} \ 4^{z_1+...+z_E}}{ \Gamma(-z_1)...\Gamma(-z_E) \Gamma(D+z_1+...+z_E)}
\ \prod_{i<j} \int_{-i\infty}^{i\infty} \frac{dw_{ij}}{2\pi i} \non\\
&& \ \ \ \ \ \ \ \ \ \ \ \
\ \Gamma(D/2+\sum_i z_i -\sum_{j \neq i} w_{ij}) \ \prod_i \Gamma(-z_i +\sum_{j \neq i} w_{ij})
\ \prod_{i<j} \ \Gamma(-w_{ij}) \left( \frac{1-X_i \cdot X_j}{2} \right)^{w_{ij}}
\eena
The parameters $z_i := z_{i(E+1)}$ are associated with the lines $(i(E+1))$ of the graph $G$.
The contours in the expression on the right side run parallel to the imaginary axis in such a manner that
the left- and right- series of poles of any gamma-function are to the left resp. right of
the integration contour. Thus,
\begin{itemize}
\item $\R(w_{ij}) < 0$ for any $i<j$,
\item $\R(-z_i + \sum_{j \neq i} w_{ij})>0$ for any $i$,
\item $\R(\sum_i z_i- \sum_{j \neq i} w_{ij}) > D/2$.
\end{itemize}
These conditions are not mutually compatible for all $z_1, ..., z_E$, and indeed the
derivation of the formula is valid only if these conditions can be satisfied.
This is the case e.g. if $-\epsilon<\R(z_i)<0$ for small $\epsilon>0$. For general
values of $z_i \in \mc$, we must analytically continue our formula. For example,
if we want to analytically continue all $z_i$ to an open neighborhood of $z_i=0$, we must move
the $z_i$ around some poles of the gamma-functions $\Gamma(-z_i +\sum_{j \neq i} w_{ij})$, and this will give
rise to corresponding residue. Hence, the resulting formula for $\M_G(X_1, \dots, X_E)$
valid for $\R(z_{ij})$ in an open neighborhood of 0 will be contain further residue
terms in addition to the terms on the right side of~\eqref{master0} such as (e.g. for $E=3$):
\bena
&&M_G(X_1, X_2, X_3) = \ \ \ \dots \ \ \ + \\
&&4\pi^{D/2+1} 2^{z_1+z_2+z_3} \ \left( \frac{1-X_1 \cdot X_2}{2} \right)^{z_1+z_2-z_3}
\left( \frac{1-X_2 \cdot X_3}{2} \right)^{z_2+z_3-z_1}
\left( \frac{1-X_3 \cdot X_1}{2} \right)^{z_3+z_1-z_2} \non\\
&&\cdot \ \frac{\Gamma(-\frac{1}{2}(z_1+z_2-z_3)) \Gamma(-\frac{1}{2}(z_2+z_3-z_1))
\Gamma(-\frac{1}{2}(z_3+z_1-z_2))}{\Gamma(-z_1) \Gamma(-z_2) \Gamma(-z_3) \Gamma(D/2 + 1/2 + z_1 + z_2 + z_3)}
 \ . \non
\eena
These additional residue terms result from moving the $z_i$ around the poles.

\subsubsection{General graph}

For a general graph, one can derive formulae of a similar nature. One possibility
is to iterate the formula for a single interaction vertex, as pointed out in~\cite{marolf2}.
The resulting formula is derived in appendix~\ref{app:D}. Another formula of this type
based on the use of graph polynomials and their special properties was derived in
sec.~5 of~\cite{self} \footnote{This paper dealt with a massive field.
However, the integrals over $X_j$ in the massless case are of precisely the same form as in the
massless case--the difference is only in the nature of the subsequent
 $z_\ell$-integrations.}. To state that formula, it is first necessary to introduce
some notation. First, we introduce a graph $G^*$ whose set of vertices consists of the vertices
of the original graph $G$, together with an additional `virtual' vertex, called $*$. The edge set of the graph $G^*$
consists of one edge $(i*)$ connecting vertex $i$ with the vertex $*$, together with one edge $(ij)$
for each factor $[(X_i-X_j)^2]^{z_{ij}}$ in eq.~\eqref{ig}, i.e. for each pair of vertices $i$ and $j$
that are connected in $G$ by at least one edge. In our formula, there will be integration parameters $w_F \in \mc$ labeled by ``forests'' $F$ within a graph $G^*$.  A ``forest'' is defined to be a subgraph
$F \subset G^*$ having the same vertices as $G$, but no loops, and a connected component of
a forest is hence a ``tree''. The forests that we consider here
have either $E$ trees or $E+1$ trees. In addition, the former
forests have precisely one tree connecting an external vertex $r \in \{1,...,E\}$ with
another external vertex $s \in \{1, ..., E\}$ or with $s=*$. An example of a graph $G$, and
a corresponding forest is drawn in the following pictures, where $E=4$.

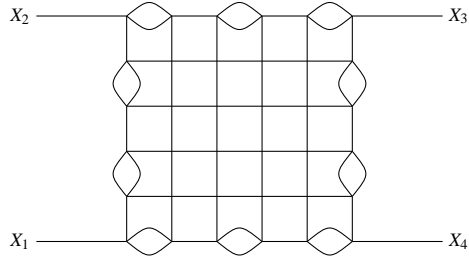
\begin{center}
\begin{tikzpicture}[scale=.6, transform shape]
\draw[black] (0,5) .. controls (.5,5.4) and (.5,5.4) .. (1,5);
\draw[black] (0,5) .. controls (.5,4.6) and (.5,4.6) .. (1,5);
\draw[black] (2,5) .. controls (2.5,5.4) and (2.5,5.4) .. (3,5);
\draw[black] (2,5) .. controls (2.5,4.6) and (2.5,4.6) .. (3,5);
\draw[black] (4,5) .. controls (4.5,5.4) and (4.5,5.4) .. (5,5);
\draw[black] (4,5) .. controls (4.5,4.6) and (4.5,4.6) .. (5,5);
\draw[black] (0,0) .. controls (.5,.4) and (.5,.4) .. (1,0);
\draw[black] (0,0) .. controls (.5,-.4) and (.5,-.4) .. (1,0);
\draw[black] (2,0) .. controls (2.5,.4) and (2.5,.4) .. (3,0);
\draw[black] (2,0) .. controls (2.5,-.4) and (2.5,-.4) .. (3,0);
\draw[black] (4,0) .. controls (4.5,.4) and (4.5,.4) .. (5,0);
\draw[black] (4,0) .. controls (4.5,-.4) and (4.5,-.4) .. (5,0);
\draw[black] (0,1) .. controls (-.4,1.5) and (-.4,1.5) .. (0,2);
\draw[black] (0,1) .. controls (.4,1.5) and (.4,1.5) .. (0,2);
\draw[black] (0,3) .. controls (-.4,3.5) and (-.4,3.5) .. (0,4);
\draw[black] (0,3) .. controls (.4,3.5) and (.4,3.5) .. (0,4);
\draw[black] (5,1) .. controls (4.6,1.5) and (4.6,1.5) .. (5,2);
\draw[black] (5,1) .. controls (5.4,1.5) and (5.4,1.5) .. (5,2);
\draw[black] (5,3) .. controls (4.6,3.5) and (4.6,3.5) .. (5,4);
\draw[black] (5,3) .. controls (5.4,3.5) and (5.4,3.5) .. (5,4);
\draw[black] (1,5) -- (2,5);
\draw[black] (3,5) -- (4,5);
\draw[black] (1,0) -- (2,0);
\draw[black] (3,0) -- (4,0);
\draw[black] (0,0) -- (0,1);
\draw[black] (0,2) -- (0,3);
\draw[black] (0,4) -- (0,5);
\draw[black] (5,0) -- (5,1);
\draw[black] (5,2) -- (5,3);
\draw[black] (5,4) -- (5,5);
\draw[black] (0,4) -- (5,4);
\draw[black] (0,3) -- (5,3);
\draw[black] (0,2) -- (5,2);
\draw[black] (0,1) -- (5,1);
\draw[black] (1,0) -- (1,5);
\draw[black] (2,0) -- (2,5);
\draw[black] (3,0) -- (3,5);
\draw[black] (4,0) -- (4,5);
\draw[black] (-2,5) node[black,left]{$X_2$} -- (0,5);
\draw[black] (-2,0) node[black,left]{$X_1$} -- (0,0);
\draw[black] (5,5) -- (7,5) node[black,right]{$X_3$};
\draw[black] (5,0) -- (7,0) node[black,right]{$X_4$};
\end{tikzpicture}
\end{center}
An associated forest \textcolor{red}{$F$} in connecting
$X_2$ with $X_4$ can look like this:
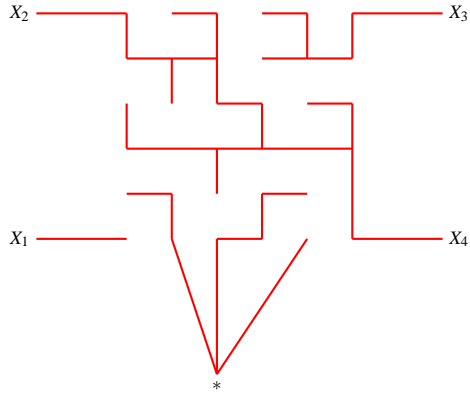
\begin{center}
\begin{tikzpicture}[scale=.6, transform shape]
\draw[red,thick] (0,2) -- (0,3);
\draw[red,thick] (0,4) -- (0,5);
\draw[red,thick] (1,0) -- (1,1);
\draw[red,thick] (1,3) -- (1,4);
\draw[red,thick] (2,1) -- (2,2);
\draw[red,thick] (2,3) -- (2,5);
\draw[red,thick] (3,0) -- (3,1);
\draw[red,thick] (3,2) -- (3,3);
\draw[red,thick] (4,4) -- (4,5);
\draw[red,thick] (5,0) -- (5,3);
\draw[red,thick] (5,4) -- (5,5);
\draw[red,thick] (-2,5) node[black,left]{$X_2$} -- (0,5);
\draw[red,thick] (1,5) -- (2,5);
\draw[red,thick] (3,5) -- (4,5);
\draw[red,thick] (0,4) -- (2,4);
\draw[red,thick] (3,4) -- (5,4);
\draw[red,thick] (2,3) -- (3,3);
\draw[red,thick] (4,3) -- (5,3);
\draw[red,thick] (0,2) -- (5,2);
\draw[red,thick] (0,1) -- (1,1);
\draw[red,thick] (3,1) -- (4,1);
\draw[red,thick] (-2,0) node[black,left]{$X_1$} -- (0,0);
\draw[red,thick] (2,0) -- (3,0);
\draw[red,thick] (5,0) -- (7,0) node[black,right]{$X_4$};
\draw[red,thick] (2,-3) node[black,below]{$*$} -- (1,0);
\draw[red,thick] (2,-3) -- (2,0);
\draw[red,thick] (2,-3) -- (4,0);
\draw[red,thick] (5,5) -- (7,5) node[black,right]{$X_3$};
\end{tikzpicture}
\end{center}

The integration variables $\vec w=\{w_F\}$ are not independent, but satisfy the constraint
\ben\label{zij1}
z_{ij} = \sum_{F \owns (ij)} w_F
\een
for any $(ij)$ such that there is an edge $\ell \in G$ connection $i$ with $j$,
which we simply write as ``$(ij) \in G$''.
We pick a subset of the variables $\vec w$
that are linearly independent from these conditions. Our choice is the
following. First, we pick, for each $(ij) \in G$, a particular forest
$F_{(ij)}$ having the property that $F_{ij} \cap G = (ij)$, so that
the remaining lines in $F_{(ij)}$ are all from the set $\{(i*) \ : \ i=E+1, ..., E+V\}$.
The corresponding variable $w_{F_{(ij)}}$ is the eliminated via eq.~\eqref{zij}.
The remaining $w_F$'s, i.e. the ones for which $F$ is neither equal to $F_{(ij)}$ for any $(ij)$,
nor equal to $\{(i*) \ : \ i=E+1, ..., E+V\}$, is denoted $\vec w$. Let us define
the meromorphic kernel $K_G$ by
\bena\label{gammag}
&&K_G (\vec w, \vec z ) = (\tfrac{1}{2} \pi^{(D+1)/2})^V \ \frac{
\Gamma(\frac{D+1}{2} + \sum_{(ij) \in G} (z_{ij} - \sum_{F \owns (ij)} w_F) + \sum_F w_F)
}
{
\Gamma(
\sum_{(ij) \in G} z_{ij} - \sum_{(ij) \notin G} \sum_{F /\!\!\!\!\!\owns (ij)} w_F)
}\\
&& \vspace{3cm} \cdot \ \frac{\ \prod_F \Gamma(-w_F) \ \prod_{(ij) \in G} \Gamma ( \sum_{F \owns (ij)} w_F - z_{ij} )}{\prod_{(ij) \notin G}  \Gamma(\frac{D+1}{2} + \sum_{F /\!\!\!\!\!\owns (ij)} w_F) \ \prod_{(ij) \in G} \Gamma(-z_{ij})} \
\wedge_F \frac{dw_F}{2\pi i} \ .\non
\eena
 All sums or products over $F$ in these expressions
 by definition exclude the forest $F=\{ (i*) \ : \ i=E+1,...,E+V \}$, and they also exclude the forests
$F_{(ij)}$, whose corresponding integration variables have been eliminated via~\eqref{zij1}. With this notation, our formula, adapted from~\cite{self}, is:
\ben\label{gk}
\boxed{
\M_{G}(X_1, \dots, X_E) =
\int_{C(\vec z)}
K_G ( \ \vec w, \vec z \ ) \
\prod_{1 \le r < s \le E} [2(1-X_r \cdot X_s)]^{\alpha_{rs}(\vec w)} \  .
}
\een
The complex number $\alpha_{rs}(\vec w)$ in the exponent is
\ben\label{alpha}
\alpha_{rs}(\vec w) = \sum_{F \ {\rm connects} \ r,s} w_F \ ,
\een
where we sum over all forests
consisting of $E$ disjoint trees, one of which is connecting the vertices $X_r$
and $X_s$, see the above picture for an example. The integrals over $\vec w$ are along a multi-dimensional
contour $C(\vec z)$ such that $\R(w_F)=\cst$ for all $F$ and such that the following conditions hold:
\begin{itemize}
\item $\R(w_F)<0$ for all $F$,
\item $\R\left[ \sum_{F \owns (ij)} w_F - z_{ij} \right]>0$ for all $(ij) \in G$,
\item $\sum_{(ij) \in G} \R\left[ \sum_{F \owns (ij)} w_F - z_{ij} \right]<(D+1)/2$.
\end{itemize}
These conditions ensure that the arguments of the gamma-functions in the numerator
of $K_G$ have positive real parts, and hence no poles. As with the alternative representation
given by eq.~\eqref{gk1}, one can show that the $\vec w$-integrals are absolutely convergent.
This follows essentially because both formulas are equivalent up to a change of integration variables.
The proof of absolute convergence of the alternative form~\eqref{gk1} is provided in appendix~\ref{app:D}.

The first condition on the integration contours is not
compatible with the other two if $\R(z_{ij})\ge 0$ for some $(ij) \in G$.
However, we must insert the integral formula~\eqref{gk} into eq.~\eqref{ig} and perform the subsequent integrations over $z_\ell$ along the contour $c$ encircling the poles at $z_\ell =\frac{D-2}{2}, ..., -1, 0$ to obtain $I_{G'}$. This means
that, $\R(z_{ij})$ will become non-negative. Thus, we have to analytically continue our integral formula~\eqref{gk} in $z_{ij}$. What happens is that the $w_F$-contours might have to be moved across some poles of the Gamma-functions in the numerator of $K_G$, and we pick up corresponding residue.

It was shown in~\cite{self} that our formula for $\tilde I_{G}$ renormalization. More
precisely, we showed that eq.~\eqref{feyn}
 can be continued analytically to a meromorphic function of the variables $z_{ij}$ in a subset of the complex plane where an ``absence of resonance condition'' is satisfied, stating that no integer linear
 combination $\sum n_{ij} \ \R(z_{ij}) \in \mz$. Eq.~\eqref{gk} gives an expression for this
 analytic continuation. To obtain $\tilde I_G$, we have to further integrate this expression over $z_\ell, \ell \in \E G$
along the collection of circles $c_\ell = c_0 \cup ... \cup c_{-D/2+1}$ [cf. eq.~\eqref{ck}]. The radii $\epsilon_\ell$ must--and can--
be chosen so that the absence of resonance condition is satisfied. Thus,
our procedure to define $I_G$ in effect involves a specific ``renormalization scheme'', and
any other scheme will lead to a different prescription for $\tilde I_G$ that is connected to the one given by
adding finite ``counterterms'' of the appropriate dimension to the action, as described in detail for curved spacetime in~\cite{hollandswald1}.
The type of possible counterterms will as usual depend on whether one has a renormalizable, or non-renormalizable interaction.

\section{Masseless deSitter quantum field theory}

\subsection{Analytic continuation}\label{analyticcont}

The deSitter manifold can be defined as the submanifold of $(D+1)$-dimensional
Minkowski space $\mr^{D+1}$ given by
\ben\label{dsdef}
dS_D = \{ X \in \mr^{D+1} \ : \  X \cdot X = -X_0^2 + \dots + X_D^2 = 1 \} \ ,
\een
with the induced metric.

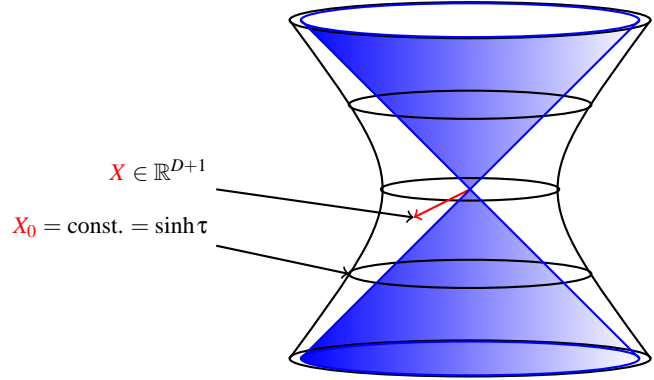
\begin{center}
\begin{tikzpicture}[scale=.75, transform shape]
\shade[left color=blue] (-3,-3) -- (0,0) -- (3,-3) -- (-3,-3);
\shade[left color=blue] (-3,3) -- (0,0) -- (3,3) -- (-3,3);
\shade[left color=blue] (0,-3) ellipse (3 and .3);
\path[fill=white] (0,3) ellipse (3 and .3);
\draw[thick] (-3.2,-3) .. controls (-1,0) and (-1,0) .. (-3.2,3);
\draw[thick] (3.2,-3) .. controls (1,0) and (1,0) .. (3.2,3);
\draw[thick,black] (0,-3) ellipse (3.2 and .32);
\draw[thick,black] (0,3) ellipse (3.2 and .32);
\draw[thick,black] (0,0) ellipse (1.58 and .2);
\draw[thick,black] (0,1.5) ellipse (2.15 and .25);
\draw[thick,black] (0,-1.5) ellipse (2.15 and .25);
\draw[->, thick] (-4.5,0) node[black, above left]{\textcolor{red}{$X$} $\in \mathbb{R}^{D+1}$} -- (-1,-.5);
\draw[->, thick] (-4.5,-1) node[black, above left]{\textcolor{red}{$X_0$} $=$ const. $= \sinh \tau$} -- (-2.15,-1.5);
\draw[->, thick,red] (0,0) -- (-1,-.5);
\draw[thick,blue] (-3,-3) -- (3,3);
\draw[thick,blue] (-3,3) -- (3,-3);
\draw[thick,blue] (0,-3) ellipse (3 and .3);
\draw[thick,blue] (0,3) ellipse (3 and .3);
\end{tikzpicture}
\end{center}
In cosmology, one is mostly interested in the subregion of deSitter spacetime
sliced by flat sections which is
covered by the coordinates $(t \in \mr, \mathbf{x} \in \mr^{D-1})$ defined by
\begin{eqnarray*}\label{cosmchart}
X_0 &=& \sinh t + \frac{1}{2} \e^{t} r^2\\
X_1 &=& \e^{t} x_1\\
&\vdots& \\
X_{D-1} &=& \e^{t} x_{D-1} \\
X_D &=& \cosh t - \frac{1}{2} \e^{t} r^2 \ .
\end{eqnarray*}
In this region, the metric takes the form
\ben\label{cosmc}
ds^2 = -dt^2 + \e^{2t} d\mathbf{x}^2 \ ,
\een
where $d\mathbf{x}^2$ is the  Euclidean flat metric on $\mr^{D-1}$. The cosmological chart covers
the half $\{X_D + X_0 < 0\}$ of $dS_D$, and its boundary is sometimes called the (a) ``cosmological horizon''.
The cosmological horizon is also equal to the boundary $\partial J^+(i^-)$ of the causal future of a point $i^-$ of $\mathscr{I}^-$. The conformal diagram for the cosmological chart is:
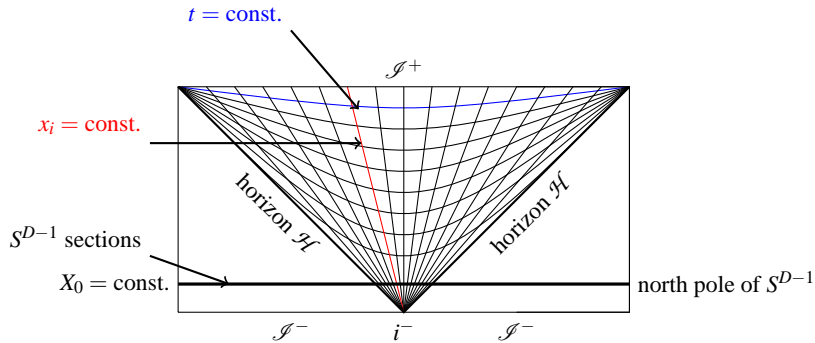
\begin{center}
\begin{tikzpicture}[scale=.75, transform shape]
\draw[blue] (-4,2) .. controls (0,1.5) and (0,1.5) .. (4,2);
\draw (-4,2) .. controls (0,1) and (0,1) .. (4,2);
\draw (-4,2) .. controls (0,0.5) and (0,0.5) .. (4,2);
\draw (-4,2) .. controls (0,0) and (0,0) .. (4,2);
\draw (-4,2) .. controls (0,-0.5) and (0,-0.5) .. (4,2);
\draw (-4,2) .. controls (0,-1) and (0,-1) .. (4,2);
\draw (-4,2) .. controls (0,-1.5) and (0,-1.5) .. (4,2);
\draw (-4,2) .. controls (0,-2) and (0,-2) .. (4,2);
\draw (0,-2) -- (0,2);
\draw (0,-2) -- (-0.5,2);
\draw[red] (0,-2) -- (-1,2);
\draw (0,-2) -- (-1.5,2);
\draw (0,-2) -- (-2,2);
\draw (0,-2) -- (-2.5,2);
\draw (0,-2) -- (-3,2);
\draw (0,-2) -- (-3.5,2);
\draw (0,-2) -- (0.5,2);
\draw (0,-2) -- (1,2);
\draw (0,-2) -- (1.5,2);
\draw (0,-2) -- (2,2);
\draw (0,-2) -- (2.5,2);
\draw (0,-2) -- (3,2);
\draw (0,-2) -- (3.5,2);
\draw[very thick] (-4,-1.5) -- (4,-1.5);
\draw[->, thick] (-4.5,-1) node[above left] {$S^{D-1}$ sections} -- (-3.1,-1.5);
\draw[->, thick] (-4.5,1) node[red, above left] {$x_i=$ const.} -- (-0.75,1);
\draw[->, thick] (-2,3) node[blue,above left] {$t=$ const.} -- (-.8,1.6);
\draw[thick] (-4,2) -- node[black,below, sloped]{horizon $\mathcal H$} (0,-2) --
node[black,below, sloped]{horizon $\mathcal H$} (4,2);
\draw (-4,-2) -- (-4,2) -- node[black,above]{${\mathscr I}^+$} (4,2) --
(4,-2) -- node[black,below]{${\mathscr I}^-$} (0,-2) node[black,below]{$i^-$} -- node[black,below]{${\mathscr I}^-$} (-4,-2);
\draw (2,-2) -- (4,-2) --  (4,2) --  (2,2);
\draw  (4,-1.5) node[black,right]{north pole of $S^{D-1}$} ;
\draw  (-4,-1.5) node[black,below,left]{$X_0=$ const.};
\end{tikzpicture}
\end{center}
In two dimensions, this conformal diagram should be thought of as the hyperboloid which has been cut
along a vertical line of constant angle. We will use the embedding coordinates $X$ in the following,
but in our final formulas one can easily go to the coordinates $(t,{\bf x})$ in the end.

As the sphere $S^D$, deSitter spacetime $dS_D$ is a constant curvature space,
but the metric is Lorentzian, rather than Riemannian. Allowing $X \in \mc^{D+1}$ in the definition~\eqref{dsdef}, one obtains a complex
manifold, $dS^\mc_D$, called {\em complex} deSitter space. The complex deSitter space
contains both the sphere $S^D=dS_D^\mc \cap (i\mr \times \mr^D)$ (i.e. taking $X_0$ purely imaginary) as well as real deSitter space $dS_D = dS^{\mc}_D \cap (\mr \times \mr^D)$ as real submanifolds. In the complex deSitter spacetime, we can introduce the $\mc$-valued ``point-pair-invariant''
\ben\label{pointpair}
Z_{12} = X_1 \cdot X_2 \ ,
\een
where the dot $\cdot$ is now the {\em Lorentzian} inner product.
The analytic continuation of the point-pair invariant $Z_{12}$ with $0$-component of
any vector $X$ taken as imaginary, $X_0 \to iX_0$, is given by the {\em Euclidean} inner product.
Therefore $|Z_{12}|\le 1$ on the sphere, but on deSitter spacetime
$|Z_{12}|$ is unbounded. The values of $Z_{12}$ in the real deSitter spacetime are closely related to the
causal relationship between $X_1,X_2 \in dS_D$. This is visualized in the following
conformal diagram of the real deSitter manifold, which indicates the values of $Z \equiv Z_{12}$
for fixed $X_2$, as $X_1$ varies:
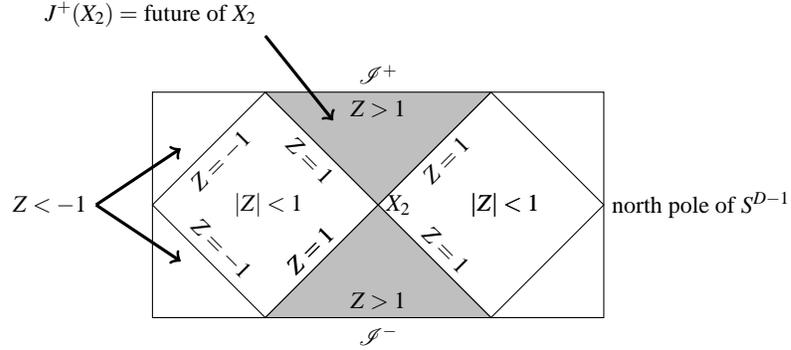
\begin{center}
\begin{tikzpicture}[scale=.75, transform shape]
\draw (0,0) -- (-2,2) -- node[black,below, sloped]{$Z=-1$} (-4,0) -- node[black,above, sloped]{$Z=-1$} (-2,-2) -- node[black,above, sloped]{$Z=1$} (0,0) node[right]{$X_2$};
\draw (-2.7,0) node[black,right]{$|Z|<1$};
\draw (3,0) node[black,left]{$|Z|<1$};
\draw (-2,-2) -- (-4,-2) --  (-4,2) -- (-2,2);
\draw (0,0) -- (2,2) -- (4,0) -- (2,-2) -- node[black,above, sloped]{$Z=1$}(0,0);
\filldraw[fill=gray!50] (0,0) -- node[black,below, sloped]{$Z=1$}(-2,2) --
node[below,black]{$Z>1$} node[above,black]{${\mathscr I}^+$} (2,2) -- node[black,below, sloped]{$Z=1$} (0,0);
\filldraw[fill=gray!50] (0,0) -- node[black,above,sloped]{$Z=1$} (-2,-2) -- node[black,above]{$Z>1$} node[below,black]{${\mathscr I}^-$}(2,-2)  --  (0,0);
\draw[->, very thick] (-2,3) node[above left] {$J^+(X_2) =$ future of $X_2$} -- (-.8,1.5);
\draw[->, very thick] (-5,0) node[left] {$Z<-1$} -- (-3.5,1);
\draw[->, very thick] (-5,0) -- (-3.5,-1);
\draw (3,0) node[black,left]{$|Z|<1$};
\draw (2,-2) -- (4,-2) -- node[black,right]{north pole of $S^{D-1}$} (4,2) -- (2,2);
\end{tikzpicture}
\end{center}
Physically, one is interested not in the correlation functions of the field
theory on $S^D$, but on $dS_D$. It is natural to conjecture that the latter can
be obtained from the former by analytical continuation through the complex deSitter space.
But it is certainly not obvious from the outset that such an analytic continuation must indeed be possible, nor
that it will give a set of correlators on the real deSitter manifold with reasonable properties.
A set of general conditions on Euclidean correlators $\langle \phi(X_1) \dots \phi(X_E) \rangle_0$
which ensure that a reasonable theory on the real deSitter can be obtained by analytic continuation
was given in~\cite{birke}, similar in spirit to the ``OS-reconstruction theorem''~\cite{osterwalder,glimm}. The key condition is that the Euclidean correlators on $S^D$ satisfy a form of ``{\em reflection positivity}''.
Unfortunately, the reconstruction theorem assumes that one has constructed the Euclidean correlators on $S^D$ {\em non-perturbatively}, whereas our construction above was essentially perturbative.
Therefore, one needs to look at this question more directly
by inspecting the analyticity properties of $\tilde I_G$, as given by eqs.~\eqref{ig} together with~\eqref{gk}, or
alternatively by~\eqref{gk1}.
These two equations tell us that $\tilde I_G(X_1, \dots, X_E)$
is a contour integral over $\vec w, \vec z$ of an expression
whose dependence on $X_1, \dots, X_E \in S^D$ enters in the combinations $(1-Z_{rs})^{\alpha_{rs}}$, where
$Z_{rs}$ is the point-pair invariant. Since each $(1-Z_{rs})^{\alpha_{rs}}$ is analytic in the cut
domain $\mc \setminus [1,\infty)$, one might expect $\tilde I_G$ to be analytic in a domain of the form
\ben
\mathscr{T}_E:=
\{ X_1, \dots, X_E \in dS^\mc_D \mid Z_{rs} \in \mc \setminus [1,\infty) \ \ \text{for $1\le r<s \le E$}\} \ .
\een
This contains many real deSitter configurations, such as configurations where all points are mutually
spacelike related to each other. To make this more precise, one has to look at the convergence
properties of integrals such as~\eqref{gk}, or~\eqref{gk1}. This is discussed briefly at
the end of appendix~\ref{app:D}. While we do not show analyticity in the entire set $\mathscr{T}_E$
there, we can show analyticity in a large subdomain. In particular, we
are able to define $I_G$ in the sense of distributions e.g. in the following situations:
\begin{enumerate}
\item[(a)] All $X_r$ are mutually spacelike related ($Z_{rs}<1$ for all $r \neq s$), or
\item[(b)]
For a fixed $r$, $X_r$ is timelike related to all other points $X_s, s \neq r$ ($Z_{rs}>1$ for all $r \neq s$)
and all points $X_s, s \neq r$ are pairwise spacelike related ($Z_{st}<1$ for all $r \neq s,t$).
\end{enumerate}
The distributional definition of $I_G$ in an open neighborhood of
such points is given by an $i\epsilon$-prescription. The correct\footnote{By ``correct'', we here
mean a prescription that gives rise to correlation functions satisfying the ``microlocal spectrum
condition'' of~\cite{brunetti}.}
$i\epsilon$-prescription is to replace the expression $(1-X_r \cdot X_s)^{\alpha_{rs}}$, in formula~\eqref{gk} (Euclidean inner product) by $(1-X_r \cdot X_s+i\epsilon s_{rs})^{\alpha_{rs}}$ (Lorentzian
inner product). $s_{rs}, r<s$
 is a sign-function,
 \ben\label{sdef}
 s_{rs} = \begin{cases}
 r-s & \text{if $X_1 \in J^+(X_2)$,}\\
 s-r & \text{if $X_1 \in J^-(X_2)$,}\\
  0 & \text{otherwise.}
 \end{cases}
 \een
We suspect that $I_G$ is in fact analytic in $\mathscr{T}_E$, but this would require a more sophisticated analysis than
that given in appendix~\ref{app:D}.

\subsection{IR-behavior}

The analytically continued $\tilde I_G$, $G$ any Feynman graph, on real deSitter is thus given by [cf. eq.~\eqref{gk}]:
\bena\label{gk2}
\tilde I_{G}(X_1, \dots, X_E) &=& \ \int_{\vec z} \ \ \prod_{\ell  \in \E G}
\frac{\Gamma(D-1+z_\ell)\Gamma(-D/2+1-z_\ell) \Gamma(z_\ell)}{ \Gamma(2 z_\ell + 1) \Gamma(-z_\ell + 1/2)}\non\\
&& \cdot \ \
\int_{C(\vec z)}
K_G ( \ \vec w, \vec z \ ) \
\prod_{1 \le r < s \le E} [2(1-Z_{rs}+i\epsilon s_{rs})]^{\alpha_{rs}(\vec w)} \  ,
\eena
where $Z_{rs}$ are the point-pair invariants in real deSitter space, and where the points $X_1, \dots, X_E$
are (for example) configurations in real deSitter space $dS_D$ described in (a) or (b) above, and not the sphere $S^D$. As we just explained, the $i\epsilon$-prescriptions means that one is dealing with a distribution, which
in general must be smeared with a suitable test-function first, after which $\epsilon>0$ is taken to zero. Compared to
to the Euclidean expressions eq.~\eqref{gk} and~\eqref{ig0}, the only difference is that
$1-X_r \cdot X_s$ (Euclidean inner product) has been replaced by $1-X_{r} \cdot X_s+i\epsilon s_{rs}$
(Lorentzian inner product), with $s_{rs}$ the sign
function given by~\eqref{sdef}.
As above,
$\vec z$ in eq.~\eqref{gk2} stands for the vector consisting of all $z_\ell, \ell \in \E G$. The $\vec z$-integral
is over the contours $z_\ell \in c$, introduced above. We obtain a completely analogous formula if the
alternative representation~\eqref{gk1} for the master integrals in app.~\ref{app:D} is used
instead. The only difference is that the kernel $K_G \to \tilde K_G$ and the exponent $\alpha_{rs} \to \tilde \alpha_{rs}$ are modified in the way given in eqs.~\eqref{tildek} resp.~\eqref{tilalpha}. Both
representations are expected to be equivalent.

If $X_r, X_s$ are time-like related (so that
$Z_{rs}>1$), the $i\epsilon$-prescription amounts to putting
\ben
[2(1-Z_{rs}+i\epsilon s_{rs})]^{\alpha_{rs}} =  \e^{is_{rs} (\pi-i\epsilon)  \alpha_{rs}} \ \left( \sinh^{2} \frac{\tau_{rs}}{2} \right)^{\alpha_{rs}} \ ,
\een
with $\tau_{rs}$ the proper time separating the points.

In typical applications in cosmology, one is interested in the equal time correlators in the cosmological chart,
$\langle \phi(t,{\bf x}_1) \cdots \phi(t,{\bf x}_E) \rangle_0$, where all point are hence pairwise spacelike related.
To obtain an expression for this, one has to simply use the expansion
of such a correlation function in terms of Feynman integrals, $\langle \prod \phi \rangle_0 = \tilde \N^{-1} \sum \lambda^{a_G} \tilde c_G \tilde I_G$, and~\eqref{gk2},
and substitute the expression for the point pair invariant $Z_{rs}$ of $(t,{\bf x}_r)$ and
$(t,{\bf x}_s)$ in the cosmological chart,
\ben
[2(1-Z_{rs}+i\epsilon s_{rs})]^{\alpha_{rs}} = \e^{2t\alpha_{rs}} |{\bf x}_r-{\bf x}_s|^{2\alpha_{rs}} \ .
\een
In particular, by taking a further Fourier transform
\ben
\hat I_G(t, \p_1; \dots ;t, \p_E) = \left( \prod_{r=1}^E \int \frac{d^{D-1} \x_r}{(2\pi)^{D-1}} \right) \e^{i\p_1 \x_1+...+i\x_E\p_E} \tilde I_G(t, \x_1; \dots; t, \x_E) \ ,
\een
in the spatial variables, eq.~\eqref{gk2}
thereby provides an expression for the contribution of the graph $G$ to the cosmological
observable eq.~\eqref{cosmobs}. This can be made somewhat more explicit using the formula for suitable $s \in \mc \setminus \mz/2, D$ even,
\bena
&&  \int d^{D-1} \x \ | \x |^{2s} \ \e^{i \p \x} \\
&=& 2^{-2s} \pi^{\frac{D-3}{2}} \ \frac{\Gamma(-s-\frac{D-3}{2}) \Gamma(2s+D-1)}{\Gamma(-2s)}
\ |\p|^{-2s-D+1} \ . \non
\eena
Then one sees that,
\bena\label{exppp}
 \hat I_{G} (t, \p_1; \dots; t,\p_E) &=&  \ \frac{\delta^{D-1}(\sum \p_i)}{a(t)^{ \ (E-1)(D-1)}} \ \int_{\vec z} \ \ \prod_{\ell  \in \E G}
\frac{\Gamma(D-1+z_\ell)\Gamma(-D/2+1-z_\ell) \Gamma(z_\ell)}{ \Gamma(2 z_\ell + 1) \Gamma(-z_\ell + 1/2)}\non\\
&& \cdot \ \
\int_{C(\vec z)}
K_G ( \ \vec w, \vec z \ ) \
F_E \left( \ \frac{\p_1}{a(t)}, \dots, \frac{\p_E}{a(t)}; \vec \alpha \ \right)  \  .
\eena
Here, $a(t) = \e^{t}$ is the scale factor in the deSitter metric $ds^2 = -dt^2 +a(t)^2 d\x^2$, $\vec \alpha$ stands for the collection $\{\alpha_{rs}(\vec w)
\ , \ r<s \}$, and $F_E$ is up to pre-factors an ordinary flat space momentum space Feynman integral:
\bena
&&\hspace{3.5cm} \ F_E ( \  \p_1, \dots , \p_E; \ \vec \alpha \ ) \\
&=&  \ \prod_{r < s}  \ \pi^{-\frac{D+1}{2}} 2^{-2\alpha_{rs}} \
\frac{\Gamma(-\alpha_{rs}-\frac{D-3}{2}) \Gamma(2\alpha_{rs}+D-1)}{ \Gamma(-2\alpha_{rs})}
\int \prod_{r < s} \ d^{D-1} \q_{rs} \ |\q_{rs}|^{-2\alpha_{rs}-D+1} \ . \non
\eena
The $\q_{rs}$-integrations are subject to the usual momentum conservation rule
\ben
\p_r = \sum_{s:s<r} \q_{sr} - \sum_{s:s>r} \q_{rs}
\een
for all $r=1,...,E$. In practice the Feynman momentum space integral $F_E$ has to be defined
carefully using a suitable analytic continuation prescription in the $\alpha_{rs}$. We will come
back to this issue in another work.

Our main aim in this section is to derive the following result:
\begin{thm}\label{thm1}
Let $r \le E$ be fixed, and let $f(X_1, ..., \hat X_r, ..., X_E)$ be a smooth function of
$E-1$ real deSitter points whose support is compact, and consists of configurations
of points $(X_1,...,\hat X_r,...,X_E)$ which are mutually spacelike.
Let $X_r$ be a point in real deSitter spacetime which is timelike to each point in the support of
$f$. Let us define
\ben
\tau := \sup \{ \tau_{rs} \ : \ s\neq r, \ \ (X_1,...,\hat X_r, ..., X_E) \in {\rm supp} \ f \}
\een
where $\tau_{rs}$ denotes the proper time between $X_r,X_s$. Then we have
\ben\label{igbound}
\Big| \int_{X_1,...,\hat X_r,...,X_E \in dS_D} \tilde I_{G}(X_1, \dots, X_E) \ f(X_1, \dots, \hat X_r, \dots, X_E) \Big| \le \cst \ \tau^N
\een
for some $N$, and a constant depending only on the graph $G$, and the function $f$.
\end{thm}
\noindent
{\em Remark:} {\em 1)} Since the correlation functions $\langle \prod \phi \rangle_0 = \tilde \N^{-1} \sum \lambda^{a_G} \tilde c_G \tilde I_G$
are sums of Feynman integrals, we get the same  growth
estimate also for the correlators. \\
{\em 2)} A look at the proof shows that the expression~\eqref{igbound} actually has an
asymptotic expansion $\sim \sum_0^N \cst_n \tau^n$ for large $\tau$.\\

\medskip
\noindent
{\em Proof:} To obtain an estimate, we first have to come to grips with the distributional nature of $\tilde I_G$.
As we have discussed in the previous subsec.~\ref{analyticcont}, $\tilde I_G$ is defined as the boundary
value of an analytic function in the neighborhood of the configurations of interest, see item (b) in subsec.~\ref{analyticcont}. The prototype of distributions of this nature are distributions
$u(x)$ on $\mr$ which are boundary values of a holomorphic function $u(x+iy)$ that is defined for
$y>0$ (and e.g. small), and which satisfy a bound of the form
\ben\label{analog}
|u(x+iy)| \le C_0 \ |y|^{-M}
\een
for some constant $C_0$, and for $x$, say, in a compact set $U \subset \mr$.
The distributional boundary value is defined in more detail as follows. Let
$z_0 = x_0+iy_0, y_0>0$ be fixed, and define, for suitable complex integration paths in the upper
half plane,
\ben
v(z) := \int_{z_0}^z dw_M \int_{z_0}^{w_M} dw_{M-1} \cdots \int_{z_0}^{w_2} dw_1 \ u(w_1) \
\een
for $z=x+iy, y>0$. Then $v$ satisfies the improved bound $|v(x+iy)| \le C_1$ for a new constant
that depends linearly on the previous constant $C_0$. Furthermore, $\partial^{M} v(x+iy) = u(x+iy)$.
Hence, if $f$ is a testfunction supported in $U$, we can define the value of the distribution $u(f)$
by the expression
\ben
u(f) := (-1)^{M} \ \lim_{\epsilon \to 0+} \int dx \ v(x+i\epsilon) \ \partial^{M} f(x)
\een
and we have the bound
\ben
|u(f)| \le C_1 \int dx \ |\partial^{M} f(x)| \ .
\een
A similar construction is possible for distributions $u$ on $U \subset \mr^n$, which are boundary
values in a region $U + iV \subset \mc^n$, where $V$ is some convex cone in $\mr^n$, and the same
type of estimates hold. Even more generally, an analog of this result holds on complex manifolds
if we pass to a local chart.

In the case at hand, the manifold in question is a set of configurations
in deSitter spacetime of the type described in (b) of subsec.~\ref{analyticcont}. By analogy with~\eqref{analog},
we are looking for a bound of the form
\ben\label{igcomplex}
|\tilde I_G(X_1, X_2 + i\epsilon e, ..., X_E + i\epsilon(E-1)e)| \le C_0 \ \epsilon^{-M} \ ,
\een
where $e \in V^+$ is in the future lightcone, and where $\epsilon>0$ is small. An estimate of this sort
is then, by analogy with the 1-dimensional case, seen to imply a distributional bound of the form
\bena
&&\Big| \int_{X_1, ...,\hat X_r,..., X_E} \tilde I_{G}(X_1, \dots, X_E) \ f(X_1, ..., \hat X_r, ..., X_E) \Big|  \\
&\le& C_1 \ \int_{X_1,..., \hat X_r, ..., X_E} |\nabla^M f(X_1, ..., \hat X_r, ..., X_E)| \ , \non
\eena
where $C_1$ depends linearly on
the previous constant $C_0$. Thus, our aim is to show that $C_0 \le \cst \ \tau^{N}$ for some $N$,
because $C_1$ will then satisfy a similar bound, hence proving the theorem.

In order to bound $\tilde I_G$ in eq.~\eqref{igcomplex}, we substitute the
representation~\eqref{ig1} in terms of the master integrals, $M_G$, for which in turn we have the
representations~\eqref{gk1} and~\eqref{bint} [or alternatively we could also use eq.~\eqref{gk2}].
This gives the estimate~\eqref{999} for the master integral, which can be stated as saying that
$C_0 \le \cst \prod_{s:s \neq r} |1-Z_{rs}|^{\sup \R(v_{rs})}$ for the analog of the above bound \eqref{igcomplex}
for the master integral. For the point-pair invariant we can use that $|Z_{rs}| \le \cst \e^\tau$, so we get the bound
$C_0 \le \cst \exp(\tau \sum \sup\R(v_{rs}))$. This implies the claim of the theorem if the suprema
$\sup \R(v_{rs})$ along the integration paths in~\eqref{bint} can be chosen {\em negative}.
Whether this is possible or not depends on the values of the complex parameters $z_\ell$ on
which the master integrals~\eqref{gk1} depend, and we now turn to this question.

According to our prescription, each $z_\ell$-integration in~\eqref{ig1} is
broken up into several small circles around the points $k=-(D-2)/2, ..., -1, 0$, cf. eq.~\eqref{ck}. Since there is one integration contour
per integration variable $z_\ell$, the $\vec z=\{z_\ell\}$-integral is a sum
\ben
\int_{\vec z} = \sum_{\vec n} \prod_{\ell \in \E G} \int_{|z_\ell - n_\ell|=\epsilon_\ell}  dz_\ell  \ ,
\een
where $\vec n=\{n_\ell\}$. The integrals that we need to look at are therefore
\bena\label{gk3}
&&\tilde I_G(X_1, X_2 + i\epsilon e, ..., X_E + i\epsilon(E-1)e) \non\\
&=& \cst \sum_{\vec n} \ \ \prod_{\ell \in \E G} \int_{|z_\ell - n_\ell|=\epsilon_\ell}  dz_\ell \
\frac{\Gamma(D-1+z_\ell)\Gamma(-D/2+1-z_\ell) \Gamma(z_\ell)}{ \Gamma(2 z_\ell + 1) \Gamma(-z_\ell + 1/2)}\non\\
&& \cdot \  M_G( \ X_1, X_2 + i\epsilon e, ..., X_E + i\epsilon(E-1)e;\ \  \{z_\ell\} \ ) \  .
\eena
The master integrals $M_G$ are given in turn by eqs.~\eqref{gk1},~\eqref{tildek} where the $\vec w$-integral is
over a multi-dimensional contour $\vec w \in \tilde C(\vec z)$. This contour is defined so that all
variables $\vec w$ in~\eqref{gk1} run parallel to the imaginary axis, and
such that any argument of a gamma function in the numerator of~\eqref{tildek} has {\em positive} real part.
Actually, as we have already discussed, such a contour may be defined only as long as all $\R(z_{ij})<0$,
which is the case if all $n_\ell <0$. Then, the contour $\tilde C(\vec z)$
can be chosen so that when $\vec w \in \tilde C(\vec z)$, then $\R(\vec w) < 0$ and
hence $\R(\tilde \alpha_{rs}(\vec w))<0$. In terms of the form~\eqref{bint} of the master integral,
this means that we can assume $\R(v_{rs})<0$ along the integration contours. Hence we get the desired decay.

The situation is more complicated when some $n_\ell$'s are
$=0$, so that some $z_\ell$'s are on the contour $|z_\ell|=\epsilon_\ell$ around $0$. Then
we can get $\R(z_\ell)\ge 0$, and consequently it may happen
that $\R(z_{ij}) \ge 0$ somewhere. In that case, the contour
$\tilde C(\vec z)$ is not defined, because one cannot achieve that all
gamma functions in the numerator of~\eqref{gk1} have arguments with positive real part
as well as $\R(\vec w) <0$ at the same time. Instead, the $\vec w$-integral
of eq.~\eqref{gk1} is now defined by analytic continuation in $\vec z$. Concretely,
this is done by moving some of the $\vec w$-contours slightly to the right across the
poles at $0$ of the gamma functions in $\tilde K_G(\vec w, \vec z)$, at the price of a corresponding residue.
Any integral that we obtain in this way is schematically of the following type:
\ben
J = \int_K \ \frac{\prod_j \Gamma( \langle a_j, \zeta \rangle + b_j)}{ \prod_j \Gamma( \langle c_j, \zeta \rangle + d_j)} \  \prod_j x_j^{\langle e_j, \zeta \rangle} \ d \zeta \ ,
\een
where $b_j, d_j \in \mz, a_j, c_j, e_j \in \mz^n$. The variable $\zeta \in \mc^n$ is a shorthand for the collection of variables $\vec z, \vec w$, and the
$x_j$'s stand for the expressions $[(1-Z_{rs})/2]$ in eq.~\eqref{gk1}.
The contour $K$ is a cartesian product of (a) small arcs $\partial \D_j$ where $\D_j$ are some discs in $\mc$, or (b) straight lines $\partial \H_j$ parallel to
the imaginary axis where $\H_j$ are some left half-planes in $\mc$.  The residue of
the integrand arise from the poles of the gamma-functions and occur if one or more of the
linear forms $\langle a_j, \zeta \rangle + b_j \in -\mn_0$ within $\D_1 \times ... \H_1 \times ...$. In terms of the
original variables $\vec w, \vec z$, this is the case by construction at most for $\vec w, \vec z$ in
eq.~\eqref{tildek} are such that a gamma function in the numerator has a pole.

To see what type of residues we can get from the integral $J$, we note a residue formula-type integral identity
$(k \ge n)$ for holomorphic $f$:
\ben
\int_{\partial \D_1 \times ... \times \partial \D_n} \frac{f(\zeta) \ d\zeta}{(\langle a_1, \zeta\rangle + b_1)...(\langle a_k, \zeta \rangle + b_k)} \  = (2\pi i)^n \ \sum_{P } k_A(P) \cdot \partial^A f(P) \ ,
\een
where the sum is over the (discrete) set of $P \in \D_1 \times ... \times  \D_n$ such that $\langle a_{j_1}, P \rangle + b_{j_1} = ... = \langle a_{j_n}, P \rangle + b_{j_n}=0$ for $n$ linearly independent linear forms.
$A\in \mn^n_0$ is a multi-index of dimension $n$, which is summed subject to the
condition that $|A|:=A_1+...+A_n \le k-n$. Furthermore, $k_A(P)$ is the `winding number'
\ben
k_A = \frac{1}{A_1! \dots A_n!} \ \ \int_{\partial \D_1 \times ... \times \partial \D_n} \frac{\ \ \langle e_1, \zeta-P\rangle^{A_1} \ ... \ \langle e_n, \zeta-P\rangle^{A_n}
}{(\langle a_1, \zeta\rangle + b_1)...(\langle a_k, \zeta \rangle + b_k)} \ d\zeta \ ,
\een
and $e_j$ the $j$-th basis vector in $\mc^n$, $\partial^A = \partial^{|A|}/\partial \zeta_1^{A_1} ... \partial \zeta_n^{A_n}$. $|A|$ is the order of the pole. Our integral formula can be used (formally) to evaluate $J$,noting that the contour $K$ is a Cartesian product of boundaries of left half-spaces (which can be thought of as infinite disks), and boundaries of discs, and noting that $1/\Gamma(z)$ is holomorphic. To make the argument rigorous,
we should approximate the half-spaces by finite discs, and in the process, we will evaluate $J$ as an infinite
sum of residues at points $P$. The convergence of this infinite sum does not have to be considered in practice,
because only finitely many residues give the dominant contribution in the large-distance analysis of~\eqref{gk3}.

In our case, $P$ consists of all possible vectors with entries $(\vec z, \vec w)$, such that a linearly independent set
of the linear conditions is satisfied which state that $(\vec z, \vec w)$ is at pole of a set of gamma functions
in the numerator of eq.~\eqref{tildek}. The number of those conditions has to be the same as the number of entries. By construction, any
such $P$ will have components with real part $\le 0$. Hence the term
with the strongest growth as $|Z_{rs}| \to \infty$ in eq.~\eqref{gk1} is one corresponding to
a residue at $P$ with all $\tilde \alpha_{rs}, z_\ell=0$. This will produce a term of the form
$\cst (\log (Z_{rs}-1))^N$, where $N$ is the order of the pole. Because $\log(Z_{rs}-1) \le \cst \tau$, his demonstrates the
claim of the theorem. \qed

\subsection{Resummation}

An unsatisfactory aspect of thm.~\ref{thm1} is that, according to the theorem, the $E$-point functions could still grow polynomially in $\tau$ for large time-like separation $\tau$.
It is conceivable that better bounds could be obtained using more refined methods.
For the 2-point function, the methods based on spectral representation
outlined in appendix~\ref{app:B} might be one possibility. The best option would of course
be to give a full non-perturbative analysis of the correlation functions, but this seems to be very difficult.
A more modest option could be to perform resummations of certain infinite classes 
of Feynman diagrams. Let us outline this here at a simple example. The action is rewritten as
\ben
I = -\lambda x^{2n} +
\int \left[\frac{1}{2} (\nabla \psi)^2 - \frac{1}{2} m^2(x) \psi^2 - \text{(higher order in $\psi$)} \right] d\mu \ ,
\een
with the usual decomposition of $\phi = x + \psi$ into the zero mode and the rest. The
quadratic term has a ``mass'' given by $m^2 = 2n(2n-1)\lambda x^2 \ge 0$. The simplest possible
resummation is to perform in closed form the perturbation series generated by this quadratic term.
As usual, carrying out the corresponding geometric series is equivalent to absorbing the mass-term into the covariance of the Gaussian measure. The new covariance is
\ben
C'(x) = C \circ \sum_{V=0}^\infty (-m^2)^V (C \circ ... \circ C)
\een
From the definition of $C$ in terms of spherical harmonics it then follows
\ben
C'(X_1, X_2;x)
= \left( \frac{1}{-\nabla^2 + m^2} \right)(X_1,X_2) -
\frac{1}{{\rm vol}(S^D)} \frac{1}{m^2} \ .
\een
Note that this covariance depends upon $x$ via $m^2 = 2n(2n-1)\lambda x^2 \ge 0$, and also note that
the difference on the right hand side gives $C'(x) \to C$ as $x \to 0$, and is therefore well-defined.
Hence, the path integral becomes
\ben
\int D\phi \ \e^{-I} \ \prod_j^E \phi(X_j) = \int_{-\infty}^{+\infty} dx \ \e^{-\lambda x^{2n}} \ \int d\nu_{C'(x)}(\psi)
\ {\rm exp}\left(-\lambda \int_{S^D} p_{\ge 3}(x, \psi) d\mu \right) \ \prod_j^E [x + \psi(X_j)] \ ,
\een
where $p_{\ge 3}$ represent the terms in $p$ that are higher than quadratic in $\psi$. The last
exponential is expanded in a Taylor series, as usual. This expansion differs from the original
perturbation expansion, because each term already involves an infinite sum of certain diagrams.

For $E=2$ points, this partially resummed path integral seems to behave better for large time-like
separation, at least to low expansion orders of the exponential. For example,
the contribution to the above path integral from the lowest order
term in the expansion of the exponential ${\rm exp}(\dots) = 1 + \dots$ is seen to
behave as $\cst \log \log Z$ for large $Z$. This behavior is better than that given by thm.~\ref{thm1}, and it shows that the large $Z$-behavior might be improved by a partial resummation. Stated differently,
the unsatisfactory growing nature of the bounds in thm.~\ref{thm1} could be an artefact due to the
truncation of the perturbation series at finite order.

A more ambitious program is to try a resummation of larger classes of diagrams, via the
so-called ``skeleton expansion'', as considered in~\cite{morrison1}, and thereby to obtain even
better bounds. However, when performing such
expansions, one has to be careful about the correct renormalization prescription, as we now briefly explain.
The point is that any change in the renormalization that we have adopted here, can be absorbed into the
addition $I \to I + \delta I$ of finite counterterms to the original action, by the general theorems~\cite{brunetti,hollandswald1,hollandswald2}. The precise form of the counterterms is dictated by
power counting and covariance. Let us focus on the case $D=2$ for simplicity, where even a non-perturbative
existence proof of the deSitter correlators is available~\cite{Jaekel}. The counterterms in $\delta I$ take the same
form as the terms already present in the action. In particular, we may get a counterterm of the form
$\delta m^2 \  \phi^2$, which is of course of the form of a mass term. Such a term will in effect make the theory
massive if ``all diagrams are summed'' i.e. in the non-perturbatively defined theory. It could lead to an exponential decay of the correlators. Of course, this is a radical change of the nature of the theory, and it should be {\em imposed} that the theory remains massless. In deSitter space, this ``renormalization condition'' is that the
support of the K\" allen-Lehmann measure $\rho(M^2) \ dM^2$ (cf. appendix~\ref{app:B}) should contain the point $M^2=0$.
Such a renormalization condition must also be respected by any resummation taking into account only a limited
class of diagrams.

\subsection{Physical consequences}\label{subsec:phys}

For a massive interacting field, the deSitter correlators $\langle \phi(X_1) ... \phi(X_E) \rangle_0$ in
the deSitter invariant (``Euclidean'') vacuum state decay exponentially in time~\cite{self,marolf1,marolf2}
to all orders in perturbation theory, but for
a massless field, we have seen that they grow polynomially in time. Although our analysis
was only graph-by-graph in a semi-perturbative setup, such a behavior, if true non-perturbatively,
can potentially have significant physical implications for the evolution of the universe on
large time-scales. To see this, let ${\mathcal O}$ be the operator
\ben
\mathcal{O} = \sum_n \ \int_{X_1,...,X_n \in dS_D} \Psi_n(X_1, ..., X_n)  \ \phi(X_1) \dots \phi(X_n)
\een
where the sum is finite and where $\Psi_n$ are some ``wave-packets''. For technical reasons--since we want to apply thm.~\ref{thm1}--we assume that the support of each $\Psi_n$ is compact, and consists of configurations $(X_1, ..., X_n)$
of points which are mutually spacelike to each other.
Let $\langle \ . \ \rangle_\Psi$ be the state obtained by applying $\mathcal{O}$ to the deSitter invariant
state $\langle \ . \ \rangle_0$, i.e.
\ben
\langle \phi(X_1) \dots \phi(X_E) \rangle_\Psi := \frac{\langle \mathcal{O}^* \ \phi(X_1) \dots \phi(X_E) \
\mathcal{O} \rangle_0}{\langle \mathcal{O}^* \mathcal{O} \rangle_0} \ ,
\een
or in ``vector notation'' (i.e. in the GNS-representation of the deSitter invariant state),
$|\Psi \rangle = \mathcal{O}|0\rangle/\| \mathcal{O}|0\rangle \|^{1/2}$. Let $\gamma(\tau)$ be a time-like curve
parameterized by proper time $\tau$, which goes to future infinity $\mathscr{I}^+$, and which
eventually becomes timelike related to any deSitter point $X_i$ in the support of $\Psi_n(X_1,...,X_n)$ for all $n$.
Then the growth of the deSitter correlators in $\langle \ . \ \rangle_0$
stated in thm.~\ref{thm1} and the following remarks immediately give:
\ben
\langle \phi[\gamma(\tau)] \rangle_\Psi \sim P(H\tau) = \cst_0 + \cst_1 (H\tau) + \cst_2 (H\tau)^2 + \dots,
\een
where we have reintroduced the Hubble constant $H$. $P$ is a polynomial, which in our
analysis, depends on the order to which the perturbation expansion is carried out.
The coefficients of the polynomial depend upon
the precise choice of the wave packets, the value of $H$, and $\lambda$. As a function
of $\lambda$, the constant term is of order one. The higher terms in the polynomial
are of order at least $\lambda^{1/2n}$ for a $\lambda \phi^{2n}$ interaction.

Although we have not analyzed composite operators in this paper, this can be done.
One obtains e.g. that, to lowest order in $\lambda$
\ben
\langle \rho[\gamma(\tau)] \rangle_\Psi  \sim \langle \rho[\gamma(\tau)] \rangle_0 + \cst \ \lambda \ (H\tau)^2 + \dots \ ,
\een
where $\rho = T_{\mu\nu}\dot \gamma^\mu \dot \gamma^\nu$ is the energy density operator evaluated
along the curve, and where $\cst$ is a constant depending on $H$ and the precise form of the
wave packets. For fixed wave packets, and small $H$, the constant would be of order $H^4$ in $D=4$.
The expectation value on the right side in the deSitter invariant state depends on the
renormalization convention for the composite operator $T_{\mu\nu}$.
Given that deSitter spacetime ought to be a solution to the semi-classical Einstein equations,
it is natural to fix the renormalization convention by requiring that
 $8\pi G \langle T_{\mu\nu} \rangle_0 = -\Lambda g_{\mu\nu}$. In $D=4$, we have $\Lambda = 3H^2$,
so we get that to lowest order in $\lambda$
\ben
\langle \rho[\gamma(\tau)] \rangle_\Psi  \sim H^2 E_P^2  \left[ \frac{3}{8\pi}  + \cst \ \lambda \left(\frac{H}{E_P}\right)^2  (H\tau)^2 + \dots
\right]
\een
with $E_P$ the Planck energy, and $\cst$ a constant of order unity.
Thus, a perturbative analysis suggests that the expected
energy density for a self-interacting, massless field in a ``typical state'' (not equal to
the deSitter invariant state) will grow in time, and could thereby
give rise in principle to significant back-reaction effects. However the linearly growing term will be comparable to the
vacuum energy term only when $\lambda (H/E_P)^2 (H\tau)^2$ is of order one, which is satisfied only when
the time-scale $\tau$ is of the order of the Hubble time, and when $\lambda > (E_P/H)^2$. Inserting the
presently observed value for $H$, this would correspond to a huge value of the coupling constant
$\lambda \gg 1$ for which our semi-perturbative analysis is clearly not applicable, and a non-perturbative
analysis will be required to settle the issue.

\vspace{2cm}

\paragraph{\bf Acknowledgements:} We have profited from discussions with I. Morrison, and are
grateful to him for providing us with a draft of his work~\cite{morrison1} which is
concerned with a similar analysis of massless deSitter quantum fields. We are also grateful
to J.~Bros, H.~Epstein, and Ch.~Kopper for discussions. This work was partly supported by
ERC grant no.~QC\&C~259562.

\appendix
\section{Spherical harmonics, Gegenbauer polynomials, K\" allen-Lehmann representation}
\label{app:B}

Here we outline how one can obtain a ``spectral representation'' of the 2-point correlation function of an interacting field analogous to the K\" allen-Lehmann representation in Minkowski spacetime. The derivation of this formula
involves spherical harmonics in $D$-dimensions, so we briefly recall their basic properties. For more
details, see e.g.~\cite{axler}. For simplicity, $H=1$ in this subsection.

\subsection{Spherical harmonics and Gegenbauer functions}

Spherical harmonics on the unit $S^D$ can be introduced via harmonic polynomials in the embedding space $\mr^{D+1}$.
A polynomial $P(X)$ on $\mr^{D+1}$ is called homogeneous of degree $h$ if $P(\lambda X) = \lambda^h P(X)$,
and it is called harmonic if it is a solution to the Laplace equation on $\mr^{D+1}$. The harmonic polynomials of
degree $h=L$ form a vector space, the dimension can be seen to be $N(L,D) =
\frac{(2L+D-1)(L+D-2)!}{(D-1)!L!}$. Spherical harmonics on $S^D$ of order $L$ are by definition just
an orthonormal basis of the space of harmonic polynomials, restricted to $S^D$. The spherical harmonics $Y_{Lj}(X), j = 1, ..., N(D,L)$ are thus normalized so that
\ben
\sum_{Lm} Y_{Lm}(X_1)^* Y_{Lm}(X_2) = \delta (X_1, X_2) \ , \quad \int_{S^D} d\mu(X) \
Y_{Lm}(X)^* Y_{L'm'}(X) = \delta_{L,L'}\delta_{m,m'}
\een
where the $\delta$ function is that on $S^D$, defined with respect to the measure $d\mu$. Expressing the Laplacian
on $\mr^{D+1}$ in polar coordinates, on sees that the spherical
harmonics are eigenfunctions of the Laplacian $\nabla^2$ on the $D$-sphere with eigenvalues $-L(L+D-1)$, so that $L$ may be viewed as the analog of the total angular momentum-, and $m$ may be viewed as the analog of the magnetic quantum numbers. One
has
\ben\label{proj}
\sum_{m=1}^{N(D,L)} Y_{Lm}(X_1)^* Y_{Lm}(X_2) = \frac{2L+D-1}{{\rm vol}(S^{D-1})} \ C_L^{(D-1)/2}(Z) \, ,
\een
where $C^{\mu}_L$ are the Gegenbauer polynomials, and where $Z$ is the point pair invariant.
The Gegenbauer polynomials are expressible in terms of a hypergeometric function,
\ben\label{gegenb}
C^{(D-1)/2}_L(Z) = \frac{\Gamma(L+D-1)}{\Gamma(D)\Gamma(L+1)} \ \hF \left( -L, L+D-1; D/2; \frac{1-Z}{2} \right) \ .
\een
Eq.~\eqref{proj} may be viewed as saying that the Gegenbauer polynomials are, up to normalization, the integral kernels
of the projector $E_L$ onto the eigenspace for the eigenvalue $-L(L+D-1)$ of the Laplacian on $S^D$. Since the dimension of this eigenspace is equal to $N(D,L)$, one gets  the orthogonality relation ${\rm tr}(E_L E_{L'})
= N(D,L) \delta_{L,L'}$. Writing out the trace of the integral kernels as integrals, one infers from this that
\ben\label{ortho}
\int_{-1}^1 dZ \ (1-Z^2)^{D/2-1} \ C^{(D-1)/2}_L(Z) C^{(D-1)/2}_{L'}(Z) = N_{D,L} \ \delta_{L,L'}
\een
for $L,L' \in \mn_0$, with normalization factor
\ben
N_{D,L} = \frac{{\rm vol}(S^{D-1})\Gamma(L+D-1)}{{\rm vol}(S^D)(2L+D-1)\Gamma(D)\Gamma(L+1)}  \ \ .
\een

\subsection{K\" allen-Lehmann measure $\rho$}

These formulas can be used to first obtain expressions for the Euclidean two-point function
for a field of mass $m^2>0$, as follows.
Writing out the condition that
$\langle \phi(X_1) \phi(X_2) \rangle_0$ is the
Euclidean Green's function of $(-\nabla^2+m^2)$ on the sphere $S^D$
gives using~\eqref{proj}
\bena\label{euclhad}
\langle \phi(X_1) \phi(X_2) \rangle_0 &=& \sum_{L,j} \frac{Y_{Lj}(X_1)^* Y_{Lj}(X_2)}{L(L+D-1)+m^2} \non\\
&=& \frac{1}{{\rm vol}(S^{D-1})} \sum_{L=0}^\infty C_L^{(D-1)/2}(Z) \frac{2L+D-1}{-c(c+D-1) + L(L-D+1)}
\eena
where $c=-(D-1)/2+[(D-1)^2/4-m^2]^{1/2}$. Using $C_L^{(D-1)/2}(Z) = (-1)^L\ C_L^{(D-1)/2}(-Z)$, and
using the above representation of the Gegenbauer polynomials as hypergeometric functions,
the above sum can be converted to a contour integral over $L$ with the help of a Watson-Sommerfeld transformation, as
observed in \cite{marolf1}:
\ben\label{euclhad1}
\langle \phi(X_1) \phi(X_2) \rangle_0 =   \int_C \frac{dL}{2\pi i} \ (2L+D-1) \ P_L \ \Delta_L(Z)
\een
where the contour $C$ is running parallel to the imaginary axis, leaving the poles in the denominator of
\ben\label{plfree}
P_L :=  \frac{1}{m^2 + L(L+D-1)}
\een
to the left, and the poles at $\mn_0$ of $\Delta_L$ to the right, see the figure.
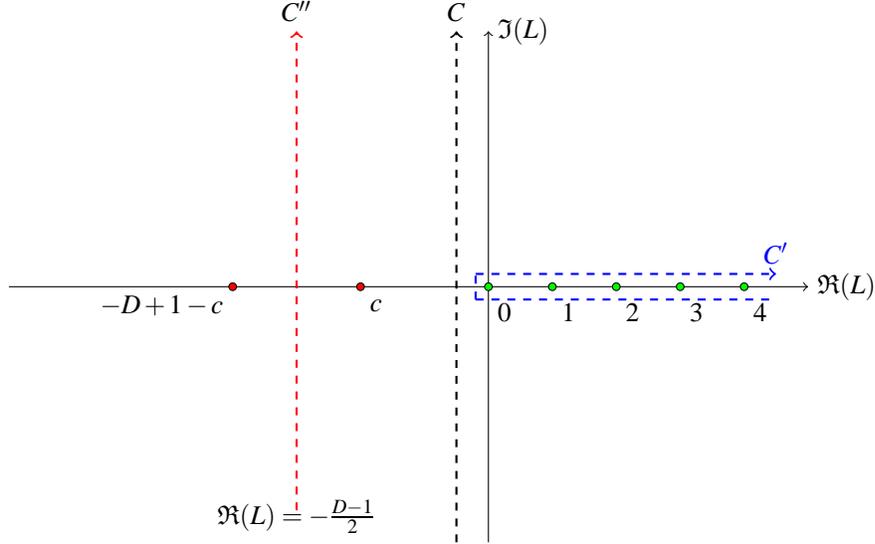
\begin{center}
\begin{tikzpicture}[scale=.85, transform shape]
\draw[->] (-7.5,0) -- (5,0) node[black,right]{$\R(L)$};
\draw (-3,-4) node[black,above]{$\R(L)=-\tfrac{D-1}{2}$};
\draw (-3,4) node[black,above]{$C''$};
\draw[->,dashed,red,thick] (-3,-3.5) -- (-3,4);
\draw (0,-0.4) node[black,right]{$0$};
\draw (1,-0.4) node[black,right]{$1$};
\draw (2,-0.4) node[black,right]{$2$};
\draw (3,-0.4) node[black,right]{$3$};
\draw (4,-0.4) node[black,right]{$4$};
\draw[->] (0,-4)  -- (0,4) node[black,right]{$\I(L)$};
\draw[->,thick,black,dashed] (-.5,-4) -- (-.5,4) node[black,above]{$C$};
\draw[->,thick,blue,dashed] (-.2,.2) -- (4.5,.2) node[blue,above]{$C'$};
\draw[thick,blue,dashed] (-.2,-.2) -- (4.5,-.2);
\draw[thick,blue,dashed] (-.2,-.2) -- (-.2,.2);
\draw (0,0) node[draw,shape=circle,scale=0.3,fill=green]{};
\draw (1,0) node[draw,shape=circle,scale=0.3,fill=green]{};
\draw (2,0) node[draw,shape=circle,scale=0.3,fill=green]{};
\draw (3,0) node[draw,shape=circle,scale=0.3,fill=green]{};
\draw (4,0) node[draw,shape=circle,scale=0.3,fill=green]{};
\draw (-2,0) node[draw,shape=circle,scale=0.3,fill=red]{};
\draw (-2,-.3) node[black,right]{$c$};
\draw (-4,0) node[draw,shape=circle,scale=0.3,fill=red]{};
\draw (-4,-.3) node[black,left]{$-D+1-c$};
\end{tikzpicture}
\end{center}
The kernel in this formula is defined as
\ben\label{spec1}
\Delta_L(Z) = \frac{1}{(4\pi)^{D/2}} \frac{\Gamma(L+D-1)\Gamma(-L)}{\Gamma(D/2)} \ \hF \left( -L, L+D-1; D/2; \frac{1+Z}{2} \right)
\een
We can deform the contour $C$ to the contour $C''$ by moving it across the pole at\footnote{Here
we assume that $m^2 \le (D-1)^2/4$, i.e. that we are in the complementary series. For the principal
series, a similar contour symmetric under $L \to -L+D-1$ is chosen.}
$L=c$. The integrand as well as the contour $C''$ is anti-symmetric under
$L \to -L+D-1$, so we are left with
the residue:
\ben\label{mprop}
\langle \phi(X_1) \phi(X_2) \rangle_0 = \Delta_c(X_1 \cdot X_2) \quad
(\equiv W(m^2; X_1, X_2))\ ,
\een
giving the free Euclidean Green's function for the mass parameter $m^2= c(c+D-1)$.
Formula~\eqref{euclhad1} can be verified as follows. The function $\Delta_L$ has poles at $L=0,1,2,...$.
Then, deforming the contour $C$ to a contour $C'$ that encircles these poles along the positive real axis
(see figure), we can
evaluate the integral by means of the residue theorem. The residue at $L =0,1,2,...$ of $(2L+D-1) \ P_L \ \Delta_L$
is precisely equal to the $L$-th term in the sum of eq.~\eqref{euclhad}.

$P_L$ is interpreted as the ``power spectrum'', or ``spectral density''. For $L \in \mn_0$, the power
spectrum can be obtained by multiplying both sides of~\eqref{euclhad} with $C_L^{(D-1)/2}(X_1 \cdot X_2)$
and integrating over $Z=X_1 \cdot X_2$ using the orthogonality of the Gegenbauer functions~\eqref{ortho}.
Using also~\eqref{spec1} and~\eqref{gegenb}, one obtains the inversion formulas:
\bena
&&P_L = \frac{\Gamma(D)}{\Gamma(-L)\Gamma(L+D-1)}
\int_{S^D \times S^D} \langle \phi(X_1) \phi(X_2) \rangle_0 \ \Delta_L(-X_1 \cdot X_2) \\
&&\langle \phi(X_1) \phi(X_2) \rangle_0 = \sum_{L=0}^\infty \frac{\sin \pi L}{\pi} \ (2L+D-1) \ P_L \ \Delta_L(-X_1 \cdot X_2) \ .
\eena
These formulas are valid not only for the free field correlators, but in fact also for the
interacting correlators (denoted in this section by $\langle \phi(X_1) \phi(X_2) \rangle_{0,\lambda}$ to
distinguish them from the free ones), because in
their derivation only the invariance property under $O(D+1)$ was used. The spectral density
in the interacting theory is denoted $P_{L,\lambda}$; of course it is no longer given by the same
formula~\eqref{plfree} as in the free theory. Thus, we have the inversion formulas
\bena\label{inversion1}
&&P_{L,\lambda} = \frac{\Gamma(D)}{\Gamma(-L)\Gamma(L+D-1)}
\int_{S^D \times S^D} \langle \phi(X_1) \phi(X_2) \rangle_{0,\lambda} \ \Delta_L(-X_1 \cdot X_2) \\
&&\langle \phi(X_1) \phi(X_2) \rangle_{0,\lambda} =
\sum_{L=0}^\infty \frac{\sin \pi L}{\pi} \ (2L+D-1) \ P_{L,\lambda} \ \Delta_L(-X_1 \cdot X_2) \ .
\eena
The first equation now has to be interpreted as the {\em definition} of $P_{L,\lambda}$.
Note that these formulas define $P_{L,\lambda}$ not only for natural numbers $L$, but
even provide an analytic continuation for complex values of $L$.
The poles in $L$ of $P_{L,\lambda}$ give information about the mass in the interacting
theory. For $\lambda = 0$, i.e. in the free theory, there are poles at $L=c$ or $L=-D+1-c$.
For a principal series scalar field, we have $c=-(D-1)/2 + i\rho, \rho \in \mr$, while
for a field in the complementary series we have $c \in [-(D-1)/2, 0]$. In an interacting
theory, the singularities of $P_{L,\lambda}$ lie in these regions, too, but they
do not have to be simple poles.

Let us suppose that we knew that there were no singularities in the region $\R(L) \ge 0$,
and let us also suppose we knew that $|P_{L,\lambda}|$ grows slower than $\e^{\pi |\I(L)|}$ for
large $|L|$. Then, because the large $|\I(L)|$ asymptotics of $\Delta_L$ is given by $O(\e^{-\pi |\I(L)|})$,
we can apply the same argument as in the free theory to get the representation
\ben\label{euclhad3}
\langle \phi(X_1) \phi(X_2) \rangle_{0,\lambda} =   \int_C \frac{dL}{2\pi i} \ (2L+D-1) \ P_{L,\lambda} \ \Delta_L(Z)
\een
also in the interacting theory. A major advantage of this formula is that
one can directly analytically continue both sides to real deSitter space; the only difference
is that on the right side, we then have to take $Z\to Z+i\epsilon s$, where $s$ is the sign that
indicates whether $X_1$ is to the future of $X_2$, or vice versa.

Eq.~\eqref{euclhad3} may be used under certain conditions to derive
an analog of the K\" allen-Lehmann representation in Minkowski spacetime, because $\Delta_L$ is
equal to the free field two-point function with $M^2 = -L(L+D-1)$. Indeed,
consider first the case that the $P_{L,\lambda}$ is analytic all the way to $\R(L)>-(D-1)/2$.
Then $C$ can be moved all the way to the left to a contour $C'$
 parallel to the imaginary axis with $\R(L) = -(D-1)/2+\epsilon$ for any $\epsilon>0$, see the
following figure.

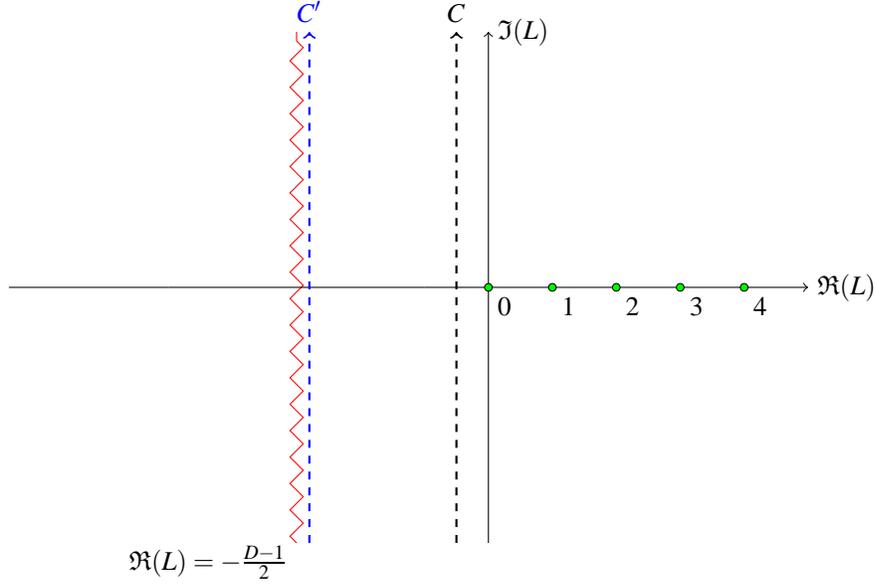
\begin{center}
\begin{tikzpicture}[scale=.85, transform shape]
\draw (-7.5,0) -- (-5,0);
\draw[->] (-1,0) -- (5,0) node[black,right]{$\R(L)$};
\draw (-3,-4.3) node[black,left]{$\R(L)=-\frac{D-1}{2}$};
\draw (0,-0.3) node[black,right]{$0$};
\draw (1,-0.3) node[black,right]{$1$};
\draw (2,-0.3) node[black,right]{$2$};
\draw (3,-0.3) node[black,right]{$3$};
\draw (4,-0.3) node[black,right]{$4$};
\draw[->] (0,-4)  -- (0,4) node[black,right]{$\I(L)$};
\draw[->,thick,black,dashed] (-.5,-4) -- (-.5,4) node[black,above]{$C$};
\draw (0,0) node[draw,shape=circle,scale=0.3,fill=green]{};
\draw (1,0) node[draw,shape=circle,scale=0.3,fill=green]{};
\draw (2,0) node[draw,shape=circle,scale=0.3,fill=green]{};
\draw (3,0) node[draw,shape=circle,scale=0.3,fill=green]{};
\draw (4,0) node[draw,shape=circle,scale=0.3,fill=green]{};
\draw[red] (-3,-4) decorate[decoration=zigzag] {--(-3,4)};
\draw (-5,0) --(-1,0);
\draw[->,thick,blue,dashed] (-2.8,-4) -- (-2.8,4) node[blue,above]{$C'$};
\end{tikzpicture}
\end{center}

The \textcolor{red}{red zigzag line} indicates the possible location of singularities such
as poles. Suppose $P_{L,\lambda}$
has a suitably distributional limit for $\epsilon \to 0$. Then using
$P_{L,\lambda}=P_{-L+D-1,\lambda}$, and changing the integration variable from
$L=-(D-1)/2 + \epsilon + i[M^2-(D-1)^2/4]^{1/2}$ to $M^2$, the above formula~\eqref{euclhad3} becomes
\ben\label{euclhad4}
\langle \phi(X_1) \phi(X_2) \rangle_{0,\lambda} =   \int_{\frac{(D-1)^2}{4}}^\infty dM^2 \ \rho_p(M^2) \ W_p(M^2;X_1, X_2) \ ,
\een
where $W_p(M^2, X_1, X_2)$ is the two-point function on deSitter spacetime of
a free field in the principal series for mass $M^2 \ge (D-1)^2/4$. The ``K\" allen-Lehmann weight'' is
given as the discontinuity
\ben\label{rhop}
\rho_p(M^2) = \frac{1}{2\pi i} \lim_{\epsilon \to 0+}
\left( P_{-\frac{D-1}{2} +  \epsilon + i[M^2-
\frac{(D-1)^2}{4}]^{1/2}, \lambda } - P_{-\frac{D-1}{2} -  \epsilon + i[M^2-
\frac{(D-1)^2}{4}]^{1/2}, \lambda} \right) \
\een
across the zigzag line.
In the general case, one expects that $P_{L,\lambda}$ can have further singularities for $L \in [-(D-1)/2,0]$. Suppose that there are no singularities in a neighborhood of $0$. Then, we may move the integration contour $C$ to the contour $C'$ depicted in the figure, where \textcolor{red}{red zigzag line} again indicates the possible location of singularities.

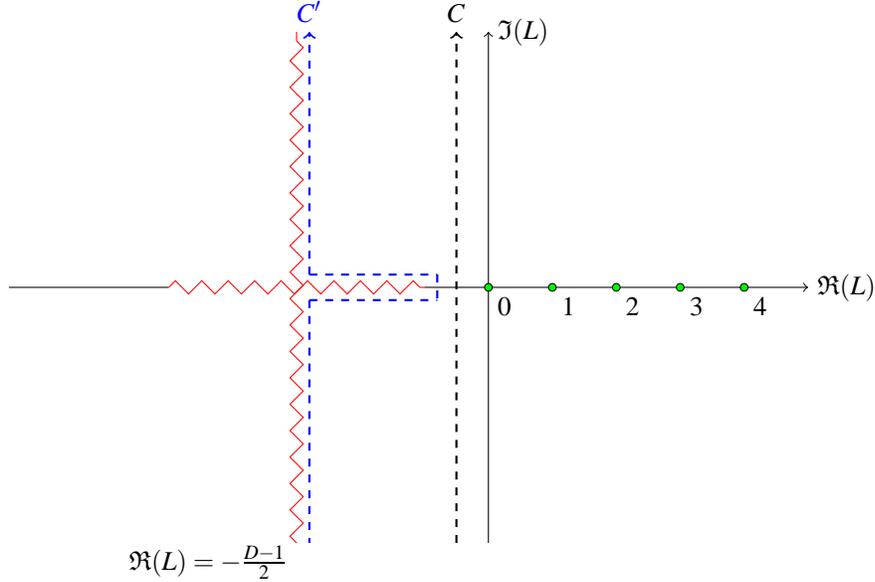
\begin{center}
\begin{tikzpicture}[scale=.85, transform shape]
\draw (-7.5,0) -- (-5,0);
\draw[->] (-1,0) -- (5,0) node[black,right]{$\R(L)$};
\draw (-3,-4.3) node[black,left]{$\R(L)=-\frac{D-1}{2}$};
\draw (0,-0.3) node[black,right]{$0$};
\draw (1,-0.3) node[black,right]{$1$};
\draw (2,-0.3) node[black,right]{$2$};
\draw (3,-0.3) node[black,right]{$3$};
\draw (4,-0.3) node[black,right]{$4$};
\draw[->] (0,-4)  -- (0,4) node[black,right]{$\I(L)$};
\draw[->,thick,black,dashed] (-.5,-4) -- (-.5,4) node[black,above]{$C$};
\draw (0,0) node[draw,shape=circle,scale=0.3,fill=green]{};
\draw (1,0) node[draw,shape=circle,scale=0.3,fill=green]{};
\draw (2,0) node[draw,shape=circle,scale=0.3,fill=green]{};
\draw (3,0) node[draw,shape=circle,scale=0.3,fill=green]{};
\draw (4,0) node[draw,shape=circle,scale=0.3,fill=green]{};
\draw[red] (-3,-4) decorate[decoration=zigzag] {--(-3,4)};
\draw[red] (-5,0) decorate[decoration=zigzag] {--(-1,0)};
\draw[->,thick,blue,dashed] (-2.8,.2) -- (-2.8,4) node[blue,above]{$C'$};
\draw[thick,blue,dashed] (-2.8,.2) -- (-.8,.2);
\draw[thick,blue,dashed] (-2.8,-.2) -- (-.8,-.2);
\draw[thick,blue,dashed] (-2.8,-.2) -- (-2.8,-4);
\draw[thick,blue,dashed] (-.8,.2) -- (-.8,-.2);
\end{tikzpicture}
\end{center}

The horizontal piece of $C'$ can be parameterized by $L=-(D-1)/2 \pm i\epsilon + [(D-1)^2/4-M^2]^{1/2}$.
Then, changing integration variable for this horizontal piece to $0\le M^2 \le (D-1)^2/4$ and assuming that the discontinuity across the zigzag line
\ben\label{rhoc}
\rho_c(M^2) = \frac{1}{2\pi i} \lim_{\epsilon \to 0+}
\left( P_{-\frac{D-1}{2} +  i\epsilon + [
\frac{(D-1)^2}{4}-M^2]^{1/2}, \lambda}-P_{-\frac{D-1}{2} -  i\epsilon + [
\frac{(D-1)^2}{4}-M^2]^{1/2}, \lambda} \right) \
\een
exists as a distribution, the K\" allen-Lehmann representation in the general case has an
additional piece and is given by
\ben\label{euclhad5}
\langle \phi(X_1) \phi(X_2) \rangle_{0,\lambda} =   \int_{\frac{(D-1)^2}{4}}^\infty dM^2 \ \rho_p(M^2) \ W_p(M^2;X_1, X_2) + \int^{\frac{(D-1)^2}{4}}_0 dM^2 \ \rho_c(M^2) \ W_c(M^2;X_1, X_2)\ .
\een
$W_c(M^2; X_1, X_2)$ is the two-point function of a free field of mass $M^2 \le (D-1)^2/4$ in the complementary series. The first integral comes from the vertical part of the contour $C'$, and
the second one from the horizontal part of the contour.
The first integral is the contribution to the K\" allen-Lehmann weight from the principal series,
the second is that from the complementary series. Formulas of this kind have previously also been derived in the pioneering work~\cite{bros}. For a free massive field of mass $m^2>0$ and $P_L$ given by
eq.~\eqref{plfree}, we get, using the standard discontinuity formula
\ben
\frac{1}{x-i\epsilon} - \frac{1}{x+i\epsilon} = 2\pi i \delta(x)
\een
that $\rho_{p,c}(M^2) = \delta(M^2-m^2)$, depending on whether $m^2$ is in
the principal- or complementary series, as is of course required.

\subsection{General consequences of K\" allen-Lehmann representation}

If $P_{L,\lambda}$ has no singularities in a neighborhood of $0$, then by definition, the K\" allen-Lehmann weight
$\rho_c$ will have its support in ${\rm supp} \rho_c \subset [m^2, (D-1)^2/2]$ for some $m^2>0$. It makes
sense to think of such theories as ``massive'', with $m^2$ the value of the mass parameter. On the other hand, if the singularities of $P_{L,\lambda}$ go all the way to $L=0$, one would speak of a ``massless theory''. In that case, the above
contour distortion argument does not work as stated, as one cannot thread a contour $C'$ between $0$
and the singular red zigzag line. However, one still expects the formula~\eqref{euclhad5} to hold true, because it simply expresses the fact that the unitary representations of $SO(D,1)$ can be decomposed
into a direct integral from representations of the principal-, complementary- and discrete series~\cite{vilenken}. Indeed, if the theory has been constructed non-perturbatively, then the positive definite property of the 2-point function,
\ben
\int_{dS_D \times dS_D} \langle \phi(X_1) \phi(X_2) \rangle_{0,\lambda} \ f(X_1) \overline{
f(X_2)
} \ge 0
\een
can be thought of as defining a positive scalar product on the space $C^\infty_0(dS_D)$
of compact support, on which the group $SO(D,1)$ acts in a natural fashion. The representation
in question is obtained from a suitable completion.
The discrete series is absent because it is in conflict with the locality
property required for the 2-point function. The connection to the representation theory
also establishes
that $\rho_{c,p}(M^2)$ are not only distributions of positive type, but that $\rho_{c,p}(M^2) \ dM^2$ are actually (positive) measures~\cite{bros}.

Let us write $\rho = 1\{M \le \frac{D-1}{2}\} \ \rho_c + 1\{M \ge \frac{D-1}{2}\} \ \rho_p$
for the combined measure on $\mr^+$ which we assume exists, and $\alpha$ be the supremum of all
numbers such that
\ben\label{assump}
\limsup_{t \to 0} \left( t^{-\alpha-1} \int_0^{t} dM^2 \ \rho(M^2) \right) < \infty \ .
\een
$\alpha$ gives a measure of the decay of the K\" allen-Lehmann weight at $M^2 = 0$.
For a massive theory, the K\" allen-Lehmann weight is supported away from $0$,
${\rm supp} \rho \subset [m^2, \infty)$, so $\alpha = \infty$. For $\alpha < 0$,
the representation eq.~\eqref{euclhad5} would be ill-defined as
$\int W(M^2; X_1, X_2) f(X_1) \overline{f(X_2)} \sim O(M^{-2})$ for small
$M^2$. For a massless theory, the topic studied in this paper, $\alpha$ is by
 definition a finite number $\ge 0$.

\begin{prop}
Suppose that a non-perturbatively defined scalar field theory has a two-point
function with K\" allen-Lehmann representation.
Suppose that K\" allen-Lehmann weight has $\alpha>0$. Then, for large time-like separation ($Z \to \infty$),
the two-point function decays as
\ben
|\langle \phi(X_1) \phi(X_2) \rangle_{0,\lambda}| \le \cst \ (\log Z)^{-\alpha} \ .
\een
\end{prop}
{\em Proof:} Of course, we use the representation $\int_0^\infty dM^2 \ \rho(M^2) \ W(M^2; X_1, X_2)$
for the 2-point function, where $W(M^2; X_1, X_2)$ is the 2-point function in the free theory
of mass $M^2$, see eq.~\eqref{mprop}. We split up the integral into
$\int_0^\infty = \int_0^{\epsilon} + \int_\epsilon^\infty$. For $M^2 \ge \epsilon > 0$, estimates on the hypergeometric
function~(see e.g.~\cite{jones}) give the uniform bound
\ben
|W(M^2; X_1, X_2)| \le \cst \ Z^{-\epsilon} \
\een
for $Z \to \infty$.
Consequently, the contribution from the second integral $\int_{\epsilon}^\infty$ is bounded by
$\cst \ Z^{-\epsilon} \le \cst (\log Z)^{-\alpha}$ for large $Z$. For $M^2 \le \epsilon$, we have instead the uniform bound
\ben
|W(M^2; X_1, X_2)| \le \frac{\cst}{M^2} \ Z^{-M^2} \ , \quad
|W'(M^2; X_1, X_2)| \le \frac{\cst}{M^4} \ Z^{-M^2} \
\een
for $Z \to \infty$, where a prime denotes a derivative w.r.t. $M^2$. Therefore,
using a partial integration together with the assumption~\eqref{assump},
\bena
\left| \int_0^\epsilon dM^2 \ \rho(M^2) \ W(M^2; X_1, X_2) \right| &=&
\left| \int_0^\epsilon W(M^2; X_1, X_2) \ d
\left( \int^{M^2}_0 d\mu^2 \ \rho(\mu^2) \right) \right| \non\\
&\le&\cst \int_0^\epsilon dM^2 \ M^{2(\alpha-1)} \ Z^{-M^2} \
\left( M^{-2(\alpha+1)} \int^{M^2}_0 d\mu^2 \ \rho(\mu^2) \right) \non\\
&\le& \cst \int_0^\infty dM^2 \ M^{2(\alpha-1)} \ Z^{-M^2} \ .
\eena
The last integral is given by $\cst (\log Z)^{-\alpha}$.
This completes the proof. \qed

The theorem gives no information when $\alpha = 0$. However, in that case we can e.g define
a refinement of this quantity by taking $\alpha$ to be the supremum over all numbers such that
\ben\label{assump1}
\limsup_{t \to 0} \left( (\log t)^{-1-\alpha} t^{-1} \int_0^{t} dM^2 \ \rho(M^2) \right) < \infty \ .
\een
Such a condition, if satisfied for $\alpha>0$, would give a decay
$|\langle \phi(X_1) \phi(X_2) \rangle_{0,\lambda}| \le \cst (\log \log Z)^{-\alpha}$.
The proof would be similar, and it is also clear that one can consider many further refinements
along those lines.

\subsection{K\" allen Lehmann measure in perturbation theory}

In perturbation theory, $\rho_{c,p}$ is only defined in perturbation theory, and we consequently cannot
apply this result directly. However, if we knew for example that
$\rho_c$ was absolutely continuous near $0$, then a qualitatively similar
definition of a quantity $\alpha$ as above can be made, such as e.g.
the supremum over all numbers such that $\limsup_{M^2 \to 0} M^{-2\alpha} \rho_c(M^2) < \infty$,
and one can derive a similar decay result.

Also in perturbation theory, we can calculate $\rho_{c,p}$ for the interacting theory as
the discontinuity of $P_{L,\lambda}$ along the zigzag line [cf. eqs.~\eqref{rhoc},~\eqref{rhop}], where $P_{L,\lambda}$ is now calculable explicitly graph-by-graph.
Indeed, breaking up the 2-point function in the first equation~\eqref{inversion1} into individual Feynman diagrams, we have, with $\tilde I_G$ the Feynman integral~\eqref{ig1}, and with
\ben\label{pgl}
P_{L,G} := \frac{\Gamma(D)}{\Gamma(-L)\Gamma(L+D-1)}
\int_{S^D \times S^D} \tilde I_G(X_1, X_2) \ \Delta_L(-X_1 \cdot X_2) \
\een
that
\ben\label{euclhad2}
P_{L,\lambda} =  \sum_{G} \tilde c_G \ \lambda^{\alpha_G}  \ P_{L,G} \ .
\een
Consequently,
\ben
\rho_c(M^2) = \frac{1}{2\pi i} \sum_{G} \tilde c_G \ \lambda^{a_G} \ \lim_{\epsilon \to 0+}
\left( P_{-\frac{D-1}{2} +  i\epsilon + [
\frac{(D-1)^2}{4}-M^2]^{1/2}, G}-P_{-\frac{D-1}{2} -  i\epsilon + [
\frac{(D-1)^2}{4}-M^2]^{1/2}, G} \right) \ .
\een
thus giving the desired breakup of $\rho_c$ into contributions from individual graphs $G$.
A similar expression holds for $\rho_p$.

Note that the integral in eq.~\eqref{pgl} simply produces a new Feynman integral $P_{L,G}=I_{G(L)}$
where the two external points $X_1, X_2$ of $G$ have been closed off with a new type of
propagator, $\Delta_L(-X_1 \cdot X_2)$, with ``mass'' $M^2 = -L(L+D-1)$.
This gives gives a new graph $G(L)$ without external
legs, with a distinguished line carrying the new propagator $\Delta_L$ instead of $C$, see the following picture.

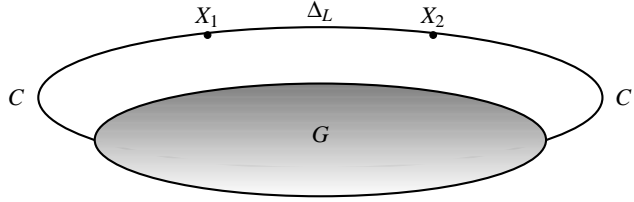
\begin{center}
\begin{tikzpicture}[scale=.75, transform shape]
\draw[thick,black] (0,.75) ellipse (5 and 1.25);
\shade[color=gray] (0,0) ellipse (4 and 1);
\draw[thick,black] (0,0) ellipse (4 and 1);
\draw (-5.1,.75) node[black, left]{$C$};
\draw (5.1,.75) node[black, right]{$C$};
\draw (-2,1.86) node[black, above]{$X_1$};
\draw (2,1.86) node[black, above]{$X_2$};
\draw (-2,1.86) node[draw,shape=circle,scale=0.3,fill=black]{};
\draw (2,1.86) node[draw,shape=circle,scale=0.3,fill=black]{};
\draw (0,1.97) node[black, above]{$\Delta_L$};
\draw (0,.4) node[black, below]{$G$};
\end{tikzpicture}
\end{center}
The new propagator can be represented in Mellin-Barnes
form as
\ben
\Delta_L(-X_1 \cdot X_2) = \frac{1}{(-\pi)^{D/2+5/2}} \int_{-\epsilon-i\infty}^{-\epsilon+i\infty} dz
\frac{\Gamma(-z)\Gamma(-L+z)\Gamma(L+D-1+z)}{\Gamma(2z+D)\Gamma(-z-D/2+1/2)}
[(X_1-X_2)^2]^z \ ,
\een
which has a similar structure as $C$, see thm.~\ref{Cthmapp}.
Therefore, in conjunction with the
Mellin-Barnes representations for $I_G$ derived in the body of the paper, this will give $P_{L,G}$
 as $I_{G(L)}$ in the form of a Mellin-Barnes integral similar to eq.~\eqref{ig}.
 One can use such representations to analyze in more detail the power spectrum $P_{L,\lambda}$ hence $\rho_{c,p}$, and to thereby analyze decay of the 2-point function. We will not do this here.

\section{Expressions for the covariance $C$}
\label{app:appa}
In this section we provide the representations for the covariance $C(X_1, X_2)$ on the sphere $S^D$
used in the main text. The discussion is somewhat different for even and odd $D$; we focus on the case when
$D$ is even, as in the main text. The definition of $C(X_1, X_2)$ is
\bena\label{Cdef}
C(X_1, X_2) &=& \sum_{L=1}^\infty \sum_{m=1}^{N(D,L)} \frac{Y_{Lm}(X_1)^* Y_{Lm}(X_2)}{L(L+D-1)} \non\\
&=& \frac{1}{{\rm vol}(S^{D-1})} \sum_{L=1}^\infty \frac{2L+D-1}{L(L+D-1)} \ C_L^{(D-1)/2}(Z) \ .
\eena
Here $Y_{Lm}, m = 1, ..., N(D,L)$ are the spherical harmonics on the unit $S^D$, and in the second
line we used the definition of the Gegenbauer functions $C_L^{(D-1)/2}(Z)$, cf. eq.~\eqref{proj}.
We also use the shorthand $Z = X_1 \cdot X_2$ as in the main text.
The above sum can be converted to a contour integral over $L$ with the help of a Watson-Sommerfeld transformation, as already explained in a similar context in the previous section:
\bena\label{inte}
C(X_1, X_2) &=& \cst \int_{-i\infty + \epsilon}^{+i\infty+\epsilon} \frac{dL}{2\pi i} \ctg(\pi L) \ \frac{2L+D-1}{L(L+D-1)} \ C_L^{(D-1)/2}(Z) \\
&=& \cst \ \int_K \frac{d\rho}{2\pi i}
 \frac{2\rho  \ \Gamma(-\rho+\frac{D-1}{2})\Gamma(\rho + \frac{D-1}{2})}{(-\rho+\frac{D-1}{2})(\rho + \frac{D-1}{2})} \non\\
&& \hspace{1cm} \times
  \ \hF \left( \frac{D-1}{2} + \rho, \frac{D-1}{2} - \rho; \frac{D}{2}; \frac{1+Z}{2} \right) \ ,
\eena
Here  $\epsilon>0$ is small.
The poles of $\ctg(\pi L)$ at $L \in \mz$ with residue $+\pi$ generate the original series
if we perform the first integral by the residue theorem, moving the contour to a contour encircling the poles of the integrand at $L \in \mz_+$. In the second line, we have used eq.~\eqref{gegenb}, the transformation of the Gegenbauer functions under $Z \to -Z$, the standard identity $\sin(\pi x) = \pi \Gamma(1+x)\Gamma(-x)$, and we have switched to the integration variable $\rho = L + (D-1)/2$. The contour $K$ is running parallel to the imaginary axis, with $\frac{D-1}{2} < \R(\rho) < \frac{D+1}{2}$.

We note the integrand in the last expression is anti-symmetric under $\rho \to -\rho$.Therefore, the
integral would vanish if the integration contour was invariant under this transformation. The contour $K$ is actually not invariant under
such a transformation, but we can deform it to such a contour $K'$, e.g. taking $K'$ to go along the imaginary $\rho$-axis.
When we deform the contour $K$ to $K'$, we cross the pole at $\rho = (D-1)/2$ of the integrand, and we pick up a corresponding
residue. To calculate that residue, we may use the standard power series expansion for $\hF$:
\bena
&&
\frac{\Gamma(-\rho+\frac{D-1}{2})\Gamma(\rho + \frac{D-1}{2})}{\Gamma(D/2)} \ \hF \left( \frac{D-1}{2} + \rho, \frac{D-1}{2} - \rho; \frac{D}{2}; \frac{1+Z}{2} \right) \non\\
&& \hspace{1cm} =   \sum_{n=0}^\infty
\frac{\Gamma(-\rho+\frac{D-1}{2}+n)\Gamma(\rho + \frac{D-1}{2}+n)}{n! \Gamma(D/2 + n)} \
\left(\frac{1+Z}{2}\right)^n \ .
\eena
As is clear from this expression, the $n=0$-term has a pole at $\rho = (D-1)/2$, giving rise to a double pole in
the integral eq.~\eqref{inte}, whereas the other $n>0$
terms are analytic there, giving rise to single poles in that integral. Thus, using the residue theorem
and keeping track of the constants, we get:
\ben
C(X_1,X_2) = \frac{1}{(4\pi)^{D/2}}
\left\{ \frac{\Gamma(D-1)}{\Gamma(D/2)} [\psi(D) - \psi(1)] + \sum_{n>0} \frac{\Gamma(D-1+n)}{n \ \Gamma(D/2 + n)} \left( \frac{1+Z}{2} \right)^{n} \right\} \ ,
\een
where $\psi = \Gamma'/\Gamma$ is the psi-function.
We can now perform the infinite sum using the series $\sum_{n>0} x^n/n = -\log(1-x)$, and
we may also use the standard identity $\psi(n)-\psi(1) = h_n$ for the psi-functions, where
\ben
h_n = 1 + \frac{1}{2} + \frac{1}{3} + \dots + \frac{1}{n-1}
\een
are the harmonic numbers. Then we easily get the following result, which is our first
alternative representation of $C$:
\begin{thm}
The covariance $C$ defined by eq.~\eqref{Cdef} is given by
\bena
&& \hspace{2cm} C(X_1, X_2) = \frac{1}{(4\pi)^{D/2}} \Bigg\{ \frac{\Gamma(D-1) \ h_D}{\Gamma(D/2)} \\
&& - \ \left( \frac{1+Z}{2} \right)^{-D/2+1} \left( \frac{\partial}{\partial Z} \right)^{D/2-1} \left[
\left( 1+Z \right)^{D-2} \log \left( \frac{1-Z}{2} \right)
\right] \Bigg\} \ . \non
\eena
\end{thm}
\noindent
We can carry out the differentiations in the last expression. If this is done, we find
\bena\label{expanded}
C(X_1, X_2) &=& \frac{1}{(4\pi)^{D/2}} \bigg\{\left( \frac{1-Z}{2} \right)^{-D/2+1} \sum_{n=0}^{D/2-2}
\frac{\Gamma(D/2+n)}{n!(D/2-1-n)} \left( \frac{1-Z}{2} \right)^{n}
+\non\\
&& \vspace{1cm} + \ \frac{\Gamma(D-1)}{\Gamma(D/2)} \left[ h_D + \psi(D-1)-\psi(D/2) + \log \left( \frac{1-Z}{2} \right) \right]
\bigg\} \ .
\eena
We claim that the sum on the right side can again be expressed via a contour integral, which is
our second alternative representation:
\begin{thm}\label{Cthmapp}
The covariance $C$ defined by eq.~\eqref{Cdef} is given by the following contour integral:
\ben
C(X_1, X_2) = \frac{h_D}{{\rm vol}(S^D)} + \frac{1}{(-4\pi)^{D/2+1/2}} \int_c dz
 \ \frac{\Gamma(D-1+z)\Gamma(-D/2+1-z)\Gamma(z)}{\Gamma(2z+1)\Gamma(-z+1/2)}
\ [(X_1-X_2)^2]^z
\een
The contour $c$ is encircling the poles at $z = -D/2+1, -D/2+2, ..., 0$, and $h_D$ are the
harmonic numbers.
\end{thm}
{\em Proof:}
Using the definition $Z=X_1 \cdot X_2$,
 the doubling identity $\Gamma(x)\Gamma(x+1/2) = 2^{-2x+1} \sqrt{\pi}\Gamma(2x)$
as well as $\Gamma(1/2+x)\Gamma(1/2-x) = \pi/\cos \pi x$, the integrand is seen to be equal to
\ben
 \ \frac{\Gamma(D-1+z)\Gamma(-D/2+1-z)}{z}
\ \cos \pi z \ \left( \frac{1-Z}{2} \right)^z
\een
up to a constant independent of $z$.
We now perform the integral by the residue theorem, using the well known residues
${\rm Res}_{z=-n} \Gamma(z) = (-1)^n/n!$. This is seen to result
precisely in the terms in expression~\eqref{expanded}. Note that the pole at $z=0$ is a double pole, and this gives
rise to the logarithm of $(1-Z)/2$ and the psi-functions. \qed

\section{Alternative form of master integrals $M_G$}\label{app:D}

For a general graph $G$, a corresponding Mellin-Barnes type formula for the master integrals $\M_G$ [cf. eq.~\eqref{feyn}] can be obtained
in different ways. One method, developed in~\cite{self}, is based on the use of graph-polynomials and was described in
the main text, leading to eq.~\eqref{gk}. Another method~\cite{marolf2} is to proceed by induction, integrating the vertices $X_{E+1}, \dots, X_{E+V} \in S^D$ in eq.~\eqref{feyn} one by one, and using at each step eq.~\eqref{master0}. We now present the result of this second method here, adapted somewhat from \cite{marolf2}. It yields a representation of the form
\ben\label{gk1}
\M_{G}(X_1, \dots, X_E) = (4\pi)^{V(D+1)/2}
\int_{\vec w}
\tilde K_G ( \ \vec w, \vec z \ ) \
\prod_{1 \le r < s \le E} \left( \frac{1-Z_{rs}}{2} \right)^{\tilde \alpha_{rs}(\vec w)} \  ,
\een
where the $\vec w$ integrations are over certain contours parallel to the imaginary axis,
and are {\em absolutely convergent}. $Z_{rs} = X_r \cdot X_s$ is the Euclidean inner product
on the sphere, which after analytic continuation to deSitter space becomes the Lorentzian
inner product.
This is of the same general form as the formula for the master integral given in the
body of the text,~\eqref{gk}, but the kernel $\tilde K_G$, the exponent $\tilde \alpha_{rs}$,
and the integration variables $\vec w$,
are defined differently here. To describe them, one first needs to introduce a somewhat elaborate notation.

The graph $G$ may have multiple lines going between a pair of vertices $i,j$. We
replace such multiple lines by a single line, which carries a corresponding complex parameter
$z_{ij}$ as in eq.~\eqref{zij}. We call the new graph $G$ as well for simplicity.
We next choose an ordering of the integration vertices $i=E+1, \dots, E+V$
which specifies the order in which the subsequent integrations of the corresponding $X_i$'s are done
in eq.~\eqref{feyn}.
For each $i=E+1, \dots, E+V$, we introduce two auxiliary graphs $C_j, G_j, j=1,...,V$ as follows.
The graph $C_j$ is a `complete graph', meaning that each vertex is joined to any other vertex by one line.
The vertex set of $C_j$ is a subset of vertices from $G$ and is obtained as follows. We first consider
the interaction vertices $\{E+V+1-j, ..., E+V\}$ of $G$, and call this set $V_j$. Then we let the
vertex set $\V C_j$ of our complete graph to be the set of all $k$ such that $(ik) \in G$ for some $i \in V_j$.
The edge set $\E C_j$ consists of all pairs $(ik)$ with $i,k \in \V C_j$, because $C_j$ is by definition
a complete graph. The graph $G_j$ is in a sense the complement of $C_j$. More precisely,
the vertex set $\V G_{j}$ consists of the vertices of $G$ minus the vertices of $V_j$, while the edge set is
$\E G_{j} = \{ (ik) \mid (ik) \in G, \ \ i,k \notin V_j\}$. For each edge $(ik)$ in $C_j$, we
introduce a complex integration parameter $w_{j,ik} \in \mc$, and we denote by $\vec w$ the set
of all such parameters as $j=1,...,V$. Our formula for $\tilde K_G$ is then
\bena\label{tildek}
&&\hspace{5cm} \tilde K_G(\vec z , \vec w) :=    \\
&&\vspace{0.5cm}\non\\
&& \cdot \ \frac{\prod_j \prod_{k \in C_{j}} \Gamma(-w_{j,k(E+V-j)}+\sum_{i\in C_{j+1}} w_{j+1,ik})
\ \prod_j \prod_{k \in G_{j}} \Gamma(-z_{k(E+V-j)}+\sum_{i \in C_{j+1}} w_{j+1,ik})}{
 \prod_j
\Gamma(-\sum_{k \in G_{j}} z_{k(E+V-j)} + \sum_{k \in C_{j}} w_{j,k(E+V-j)} +D)} \non\\
&&\cdot \ \frac{\Gamma[ \
\sum_{k \in G_{j}} (-z_{k(E+V-j)} + \tfrac{1}{2}\sum_{i \in C_{j+1}} w_{j+1,ik})
+
\sum_{k \in C_{j}} (w_{j,k(E+V-j)}-\tfrac{1}{2} \sum_{i \in C_{j+1}} w_{j+1,ik})
+
D/2
 \ ]}{\prod_{i,j} \Gamma(-z_{ij}) }   \non\\
 && \hspace{4cm}
 \cdot \ 4^{\sum z_{ij}} \ {\textstyle \prod_{j>0}  \prod_{i,k=1,...,E: \ (ik) \in C_j}} \ \Gamma(-w_{j,ik}) \
 \frac{dw_{j,ik}}{2\pi i} \ . \non
\eena
The sums/products over $j$ run from $0$ to $V$ unless stated otherwise, and for $j=0$, $C_0$ is defined to consist
of the lines $((E+V)k)$, where $k$ are vertices in $G$ connected to the vertex $E+V$. For those, we are using the notation $w_{0,(E+V)k} = z_{(E+V)k}$.
Our formula for $\tilde \alpha_{rs}$ is
\ben\label{tilalpha}
\tilde \alpha_{rs}(\vec w) := \sum_{j>0: \ C_j \owns (rs)} w_{j,rs} \ .
\een
The integration over $\vec w$ is over contours parallel to the imaginary axis,
such that the real part of any argument
of a gamma-function in the numerator is $>0$.
Such contours do not exist for all choices of $z_{ij}$, but e.g. for those with $-\epsilon<\R(z_{ij})<0$
for sufficiently small $\epsilon$. For other choices, the master integral is defined by analytic continuation.
The $\vec w$-integrals are absolutely convergent.

For example, for $V=1$ ($G$ the `star graph'),
the complete graph $C_1$ consists of all edges $\{(ik) \ : \ 1 \le i<k \le E\}$,
and the kernel becomes
\ben
\tilde K_G( \vec z, \vec w) = 4^{\sum z_i} \ (4\pi)^{(D+1)/2}
\frac{ \Gamma(D/2+\sum_i z_i -\sum_{j \neq i} w_{ij}) \ \prod_i \Gamma(-z_i +\sum_{j \neq i} w_{ij})
\ \prod_{i<j} \ \Gamma(-w_{ij}) }{
\Gamma(z_1+...+z_E+D) \Gamma(-z_1) ... \Gamma(-z_E)
} \ ,
\een
in accordance with eq.~\eqref{master0}.

The representation~\eqref{gk1}, and eq.~\eqref{tildek} can be proved by induction on the number $V$ of interaction vertices of the graph.
Each time a new interaction vertex is added, we may use eq.~\eqref{master0}. The inductive process
is very similar to that described in~\cite{marolf2}. In the course of the inductive argument, one also proves that the integrals over $\vec w$ are absolutely convergent. In more detail,
if the graph $G$ has $V$ vertices, it is convenient to rename the `last' set of integration variables
$v_{rs}:=w_{V,rs}, r,s = 1,...,E$, and to define the kernel $A_G(\vec z, \vec v)$ to be
the kernel $\tilde K_G(\vec z, \vec w)$, divided by $\prod_{r<s} \Gamma(-v_{rs})$, and with all the integrations
over $w_{j,rs}, j<V$ carried out already. The master integral $M_G$ is then, after shifting
the integration variables:
\ben\label{bint}
M_G(X_1, \dots, X_E) = \cst_G \ \int_{\vec v} \ \ A_G( \ \vec z, \vec v \ ) \ \prod_{1 \le r<s \le E}
\Gamma(-v_{rs}) \ \left( \frac{1-Z_{rs}}{2} \right)^{v_{rs}}
\een
The induction hypothesis may then be formulated in terms of this
kernel. They are I1) that $A_G(\vec z, \vec v)$ is
analytic in $\vec v$ in a strip $-\epsilon < \I(v_{rs}) < 0$ [where the
integration contours run in the previous formula], and I2), that
\ben\label{AGest}
\bigg| \ A_G(\vec z, \vec v) \ \bigg| \le \cst \bigg[1+\sum_{1 \le r < s \le E} |\I(v_{rs})| \bigg]^N \  \exp\bigg(
-\frac{\pi}{2} \sum_{r=1}^E \bigg| \sum_{s \neq r} \I(v_{rs}) \bigg|
\bigg) \ ,
\een
for some $N$, and
for  sufficiently small\footnote{Note that $M_G$ is only needed for such $z_{ij}$, as the variables $z_{ij}$ are related to $z_\ell$ by eq.~\eqref{zij}, and each $z_\ell$
is integrated over a contour $c$ around $0,-1,...,-D/2+1$ which may have arbitrarily small
imaginary part.} $|\I(z_{ij})|$. This condition, together with the growth
estimate for the gamma-function, $|\Gamma(-v_{rs})| \sim \cst |\I(v_{rs})|^{-1/2+\R(v_{rs})} \exp[-\frac{\pi}{2} |\I(v_{rs})|]$ for large $|\I(v_{rs})| \to \infty$, is immediately seen to imply the absolute convergence of the
integrals over $\vec v$ in eq.~\eqref{bint} for example if all the points $X_i$ are in mutually spacelike
position, i.e. $Z_{rs}<1$ for all $r<s$. More general configurations are discussed below.

One has:

\begin{thm}
The master integral $M_G$ is given by eq.~\eqref{gk1}, with kernel $\tilde K_G(\vec z, \vec w)$
given by eq.~\eqref{tildek}, and $\tilde \alpha_{rs}$ by~\eqref{tilalpha}. The multiple integrals over $\vec w$ are along contours parallel
to the imaginary axis
leaving the left resp. right poles of any gamma-function in the numerator to the left resp. right.
Such contours exist e.g. for $-\epsilon < \R(z_{ij}) < 0$, and the corresponding integrals over
$\vec w$ are then absolutely convergent for mutually spacelike configurations of points, i.e. $Z_{rs}<1$
for all $r<s$,
at least when $|\I(z_{ij})|<\epsilon$. For general
values $\R(z_{ij})$ and $|\I(z_{ij})|<\epsilon$, $M_G$ can be
analytically continued to a (meromorphic) function.
\end{thm}

{\em Proof:} The proof is by induction in $V$ using the inductive assumptions I1) and I2), and
using at each step eq.~\eqref{master0}. The details are given \cite{marolf2}; a
difference is that our inductive bound I2) is sharper than that used in~\cite{marolf2}.
Another difference is that we give an explicit formula for $\tilde K_G$.
The bound~\eqref{AGest} is seen to reproduce itself using the estimate on the gamma-function
in the induction step, and it is satisfied for the integral~\eqref{master0}. \qed

\vspace{.5cm}
\noindent
In order to define $M_G$ for more general configurations, one has to investigate at the convergence properties of the
integrals~\eqref{gk1} or~\eqref{bint}. Assume we have a complex deSitter configuration $(X_1, \dots, X_E)$
in $\mathscr{T}_E$, i.e. $Z_{rs} \in \mc \setminus [1,\infty)$, and let $\varphi_{rs} := {\rm Arg}(1-Z_{rs}) \in
(-\pi,\pi)$. In view of $|(1-Z_{rs})^{v_{rs}}| = |1-Z_{rs}|^{\R(v_{rs})} \ \e^{\varphi_{rs}  \I(v_{rs})}$, and
of eq.~\eqref{AGest} and~\eqref{bint}, we need to look at the convergence of the integral on the right side of
\bena\label{bint1}
&&|M_G(X_1, \dots, X_E)| \le \cst \ \prod_{1 \le r<s \le E} |1-Z_{rs}|^{\sup \ \R(v_{rs})} \\
&& \ \cdot \  \int_{\vec v}
\bigg[1+\sum_{1 \le r < s \le E} |\I(v_{rs})| \bigg]^N   \exp\bigg(
-\frac{\pi}{2} \sum_{r=1}^E \bigg| \sum_{s \neq r} \I(v_{rs}) \bigg| \bigg)
\ \prod_{1 \le r<s \le E} |\Gamma(-v_{rs})| \ \e^{\varphi_{rs} \I(v_{rs})} \ . \non
\eena
The gamma functions are estimated as usual by
 $|\Gamma(-v_{rs})| \sim \cst |\I(v_{rs})|^{-1/2+\R(v_{rs})} \exp[-\frac{\pi}{2} |\I(v_{rs})|]$ for large $|\I(v_{rs})| \to \infty$.
Then, we have absolute convergence automatically if each $|\varphi_{rs}|<\pi/2$, even
without using the ``exp'' factor under this integral. This includes any
real configuration $(X_1, \dots, X_E)$ where all points are mutually spacelike to each other, i.e.
$Z_{rs}<1$ for all $r\neq s$. A slightly more careful look at \eqref{bint1} reveals that we have absolute convergence also e.g. for the following configuration needed in the main text. Let $r \le E$ be fixed, and let us assume
that $X_r$ is real and timelike w.r.t. to $X_s$ for all $s \neq r$, and that the $X_s$ are real and mutually spacelike related
for $s \neq r$. Consider instead of $(X_1,
\ldots, X_E)$ the complex configuration $(X_1, ..., X_r + i(r-1)e\epsilon, ..., X_E+i(E-1)e\epsilon)$, where $e \in \mr^{D+1}$
is some time-like vector and $\epsilon>0$ small. Then for the complex configuration $|\pi-\varphi_{rs}| \sim \cst \epsilon, r\neq s$, and $|\varphi_{st}| \sim \cst \epsilon, r \neq s,t$. The bound~\eqref{bint1} is now seen to give
\ben\label{999}
| \ M_G(X_1, ..., X_r + i(r-1)e\epsilon, ..., X_E+i(E-1)e\epsilon) \ | \le \cst \ \prod_{s:s\neq r}
|1-Z_{rs}|^{\R(v_{rs})} \ \epsilon^{-M} \ ,
\een
for some $M$,
this time using the ``exp'' factor to get the bound. Bounds of this kind, together with the analyticity of $M_G$,
imply using standard results in distribution theory (see e.g. ch.~IX of~\cite{hor}, and also the proof of~thm.~\ref{thm1} for details)
that $M_G$ is a distribution in an open neighborhood of
the real deSitter configuration $(X_1, \ldots, X_E)$.

\end{document}